\documentclass{article}

\usepackage{PRIMEarxiv}

\usepackage[utf8]{inputenc} 
\usepackage[T1]{fontenc}    
\usepackage{hyperref}       
\usepackage{url}            
\usepackage{booktabs}       
\usepackage{amsmath}
\usepackage{amsthm}
\usepackage{amssymb}
\usepackage{mathtools}
\usepackage{amsfonts}       
\usepackage{nicefrac}       
\usepackage{microtype}      
\usepackage{fancyhdr}       
\usepackage{graphicx}       
\usepackage{times}
\usepackage{subfig}
\usepackage{xcolor}
\usepackage{array}
\usepackage{cite}
\usepackage{multirow}
\usepackage{adjustbox}
\usepackage{fontawesome}
\usepackage[font=footnotesize,labelfont=bf]{caption}
\usepackage{tikz-network}
\usepackage{enumitem}

\DeclareMathOperator{\diag}{diag}
\DeclareMathOperator{\tr}{tr}

\newtheorem{theorem}{Theorem}
\newtheorem{lemma}{Lemma}
\newtheorem{proposition}{Proposition}
\newtheorem{corollary}{Corollary}

\newtheorem{assumption}{Assumption}

\newtheorem{remark}{Remark}

\title{
Input-to-State Stability in Distributed Adaptive Identification
of Nonlinear Systems without \\ Input Sharing
}

\author{
  Moh Kamalul Wafi\\
  Department of Electrical and Computer Engineering \\
  Northeastern University \\
  Boston, MA, USA\\
  \texttt{wafi.m@northeastern.edu} \\
}

\begin{document}
\maketitle

\begin{abstract}
This paper studies continuous-time distributed adaptive identification
of unknown nonlinear systems over directed communication networks
without requiring source or neighboring inputs to be shared. Each
sensing node reconstructs the source state while adapting local
estimates of the unknown linear and nonlinear dynamics using only
locally available information and neighboring state estimates. A
network-level error representation is derived to characterize the
interaction among the communication topology, observer dynamics,
parameter uncertainty, and nonlinear mismatch. Norm-based, structured
Lyapunov, and robust LMI conditions are established to guarantee that
the resulting coupling operator is Hurwitz. Unlike the corresponding
discrete-time formulation, the proposed continuous-time adaptive laws
do not require normalization. A Lyapunov analysis establishes
boundedness of the adaptive parameters and asymptotic convergence of
the estimation and source-tracking errors in the unperturbed case.
Input-to-state stability is further established under bounded predictor
mismatch and source disturbances. Numerical studies over star, cyclic,
and path topologies support the theoretical results and demonstrate
approximately linear computational scaling with the number of sensing
nodes.
\end{abstract}

\allowdisplaybreaks



\section{Introduction}
\label{sec:introduction}

Distributed sensing networks are increasingly used to monitor
dynamic processes whose states cannot be reliably reconstructed by a single
sensor. Their development builds on advances in consensus, cooperative
control, and networked multi-agent systems, which provide systematic tools for
coordinating information over directed and time-varying graphs
\cite{OlfatiSaber2004Consensus,OlfatiSaber2007Consensus,Wafi-ASCC26-FTC,Ren2005Consensus,Dorri2018Survey}.
Applications include autonomous systems, infrastructure monitoring,
industrial processes, environmental sensing, and biological networks \cite{Cao2013Overview,Wafi-CoDIT,Wafi-CoDIT-arXiv,Jadbabaie2003Coordination}.
In these settings, multiple sensing nodes combine locally available information
with data received from neighboring nodes to estimate the behavior of a
common dynamic source. Compared with centralized architectures, distributed
estimation can improve spatial coverage, scalability, and resilience while
reducing reliance on a single processing unit \cite{Moreau2005Stability,Wafi-LCSS24}.

Distributed state estimation has consequently received considerable
attention. Consensus-based Kalman filters and diffusion strategies distribute
the filtering task across sensor networks
\cite{OlfatiSaber2007DKF,Das2015Dynamic,Das2017Innovations,Cattivelli2010Kalman,Talebi2019Kalman}.
For deterministic linear systems, distributed observers have been developed
under various detectability, communication, and convergence requirements
\cite{Park2017Omniscience,Mitra2018Observers,Wang2018Observer,Han2019Simple,Ortega2021FiniteTime}.
These contributions establish important foundations, but most presume that
the monitored dynamics are known when the observer is constructed.

A more difficult problem arises when the source dynamics are unknown.
Adaptive estimation offers an online alternative to fixed-model observers and
offline data-driven training. Classical adaptive-observer designs jointly
estimate states and unknown parameters for structured nonlinear systems
\cite{Bastin1988Stable,Marino1990Adaptive,Marino1992Global,Cho1997Systematic,Wafi-Elham}.
At the network level, diffusion least-mean-square, recursive least-squares,
and related learning strategies exploit cooperation among spatially
distributed data sources
\cite{Lopes2008Diffusion,Cattivelli2008RLS,Mateos2012RLS,Sayed2013Diffusion,Sayed2014Adaptation}.
Distributed adaptive protocols have also been used to coordinate agents with
uncertain linear or Lipschitz nonlinear dynamics
\cite{Li2013Adaptive}. Nevertheless, jointly reconstructing an
unknown nonlinear source and adapting its dynamics remains challenging
because the model estimates, state estimates, and graph coupling evolve
simultaneously.

Unknown source inputs introduce a further difficulty. Classical unknown-input
observers seek to decouple or reconstruct unmeasured exogenous signals
\cite{Hou1992Unknown,Darouach1994FullOrder,Chakrabarty2017Unknown,Veluvolu2009Sliding}.
In a sensing network, however, requiring the source input to be available or
transmitted to every node may be impractical because of limited bandwidth,
privacy constraints, communication failures, or the absence of direct input
measurements \cite{Wafi-ASCC26-DNC}. Exchanging the adaptive inputs generated by neighboring nodes
similarly increases communication and storage requirements. A genuinely
distributed method should therefore reconstruct the source state and adapt
its unknown dynamics without sharing either the source input or neighboring
inputs \cite{Wafi-ASME-LDSC26}.

The interaction between adaptation and network coupling makes the stability
analysis nontrivial \cite{Wafi-ACC26-DAC}. Stability of the individual observer matrices alone does
not guarantee stability of the complete estimation network because the graph
weights, source dynamics, and local observer dynamics jointly determine the
network error operator. Nonlinear terms create an additional mismatch because
each node evaluates the nonlinear mapping at a locally aggregated state rather
than at the unavailable source state. Moreover, time-varying unknown inputs
and external disturbances generally preclude exact convergence and require
robustness estimates and small-gain arguments \cite{Teel1996SmallGain}.
Related adaptive and consensus-based diagnostic methods
illustrate both the value and the difficulty of handling uncertainty in
networked nonlinear systems \cite{Wafi-SIAM,Franco2006Diagnosis,Ferrari2012Fault}.

Continuous-time systems deserve particular attention because they naturally
describe many physical, biological, and engineering processes. Continuous-
time adaptation also has a useful structural feature: the Lyapunov derivative
is linear in the adaptive update rates. The predictor-mismatch terms can
therefore be canceled directly, without the quadratic adaptive-increment terms
that arise in a discrete-time Lyapunov difference. This permits unnormalized
adaptive laws and yields a simpler closed-loop analysis. At the same time,
local implementability requires more than the existence of an arbitrary
positive-definite Lyapunov matrix: its network-level factor must preserve a
block-diagonal structure so that each adaptive direction can be computed from
local and neighboring errors.

Motivated by these considerations, this paper develops a continuous-time
distributed adaptive identification framework for unknown nonlinear source
systems over directed communication networks. Each sensing node forms an
aggregated state from locally available source and neighboring information,
updates local estimates of the unknown linear and nonlinear dynamics, and
adjusts its own input without accessing the unknown source input or the inputs
generated by neighboring nodes. The resulting framework simultaneously
addresses state reconstruction, online model adaptation, limited
communication, local implementability, and robustness to bounded
perturbations.

The main contributions are summarized as follows:

\begin{enumerate}[leftmargin=*]

\item A continuous-time distributed adaptive identification framework is
developed for nonlinear source systems with unknown linear dynamics, unknown
nonlinear coefficients, and an unavailable source input. Each sensing node
uses only locally available information and neighboring state and error
signals, without requiring the source input or neighboring adaptive inputs to
be shared.

\item A compact network-level error model is derived to separate the effects
of the communication topology, local observer dynamics, adaptive predictor
mismatch, nonlinear mapping mismatch, and external disturbances. This
representation provides the basis for the subsequent stability and robustness
analysis.

\item Proposition~\ref{prop:NL:H_hurwitz} establishes a direct
logarithmic-norm condition for the Hurwitz stability of the network coupling
operator. Lemma~\ref{lem:NL:LMI_H} provides a structure-preserving Lyapunov
certificate with a diagonal network-level factor, while
Corollary~\ref{cor:NL:robust_H} extends the certificate to norm-bounded
uncertainty in the unknown source matrix.

\item Locally implementable, unnormalized continuous-time adaptive laws are
introduced for the unknown linear matrix, nonlinear coefficient matrix, and
local input. The structured Lyapunov matrix makes the network-weighted adaptive
direction computable from local and neighboring errors. Unlike the
discrete-time formulation, normalization is unnecessary because the adaptive
predictor mismatch is canceled exactly in the Lyapunov derivative.

\item Theorem~\ref{thm:AdpId:CL} establishes a general Lyapunov bound and, in
the nominal case, boundedness of the adaptive variables and asymptotic
convergence of the estimation and source-tracking errors under the proposed
Hurwitz and nonlinear small-gain conditions.
Corollary~\ref{cor:AdpId:ISS} provides a finite-time ISS estimate and the
corresponding asymptotic bounds under bounded predictor mismatch and source
disturbances.

\item Numerical studies over star, cyclic, and path communication topologies
validate the theoretical results. The nominal experiments demonstrate
asymptotic state reconstruction, while the perturbed experiments show bounded
tracking under a time-varying unknown input and external disturbance. The
proposed Hurwitz certificates are verified numerically, and a fixed-step
complexity benchmark demonstrates approximately linear growth in online
computational cost with the number of sensing nodes.

\end{enumerate}

The remainder of this paper is organized as follows.
Section~\ref{sec:ComNetwork} introduces the directed communication network and
its associated matrices. Section~\ref{sec:ProblemFormulation} formulates the
unknown nonlinear source and distributed adaptive observers.
Section~\ref{sec:AdaptiveId} derives the network error dynamics, establishes
the Hurwitz stability conditions, and presents the adaptive stability and
robustness results. Section~\ref{sec:NR} reports the numerical studies and
computational benchmarks. Section~\ref{sec:Conclusion} concludes the paper.

\textbf{Notation.}
$\mathbb{R}^n$ and $\mathbb{R}^{n\times p}$ denote the spaces of real
$n$-vectors and $n\times p$ matrices, respectively. The $p\times p$ identity
matrix is $I_p$. For matrices $A_i\in\mathbb{R}^{n\times q}$,
$i\in\{1,\dots,p\}$, the notation
$\diag\{A_1,\dots,A_p\}\in\mathbb{R}^{pn\times pq}$ denotes the corresponding
block-diagonal matrix. The vectors of ones and zeros in $\mathbb{R}^p$ are
$\mathbf 1_p$ and $\mathbf 0_p$, respectively. The symbols $A\otimes B$,
$\tr(A)$, $\lambda(A)$, and $\operatorname{vec}(A)$ denote the Kronecker
product, trace, spectrum, and vectorization of $A$, respectively.

\section{Communication Network}\label{sec:ComNetwork}

Consider a network of $m{+}1$ agents indexed by
$\mathcal{V}=\{0,1,\dots,m\}$. Agent $0$ represents the dynamic source
whose state is to be reconstructed and whose dynamics are unknown,
whereas agents $1,\dots,m$ constitute the sensing network. Each sensing
node operates as a \emph{sensor--observer}: it gathers locally available
information, updates its adaptive observer, and exchanges state estimates
with its in-neighbors to cooperatively reconstruct the source state while
adapting local estimates of its unknown dynamics. The source does not
participate in the distributed estimation process and solely generates
the state trajectory of interest.

The communication architecture is described by a weighted directed
graph $\mathcal{G}=(\mathcal{V},\mathcal{E},\mathcal{W})$, where
$\mathcal{E}\subseteq\mathcal{V}\times\mathcal{V}$ is the edge set and
$\mathcal{W}=\{w_{ij}\}$ is the collection of nonnegative communication
weights. An edge $(i,j)\in\mathcal{E}$ indicates that agent $i$ receives
information from agent $j$ with weight $w_{ij}>0$; otherwise,
$w_{ij}=0$. Accordingly, the in-neighborhood of agent $i$ is defined as
$\mathcal{N}_i\coloneqq\{\,j\in\mathcal{V}\mid(i,j)\in\mathcal{E}\,\}$.
The weights $\{w_{ij}\}$ are design parameters of the distributed identification scheme and do not represent physical signal strengths.

\begin{figure}[h!]
    \centering
    \scalebox{0.90}{{\begin{tikzpicture}
            \centering
            \Text[x=0,y=1.3,fontsize=\small]{$\mathcal{G}$};
            \Vertex[x=-.375,y=.375,label=$0$,color=red,opacity=0.1,size=.5]{L}
            \Vertex[x=1,y=1,label=$1$,color=green,opacity=0.1,size=.5]{1}
            \Vertex[x=-1,y=1,label=$2$,color=green,opacity=0.1,size=.5]{2}
            \Vertex[x=-1,y=-1,label=$3$,color=green,opacity=0.1,size=.5]{3}
            \Vertex[x=0,y=-1,label=$4$,color=green,opacity=0.1,size=.5]{4}
            \Vertex[x=1,y=0,label=$5$,color=green,opacity=0.1,size=.5]{5}
            \Edge[Direct,color=red,label=$w_{10}$](L)(1)
            \Edge[Direct,color=red,label=$w_{40}$](L)(4)
            \Edge[Direct,color=red,label=$w_{50}$](L)(5)
            \Edge[](1)(2)
            \Edge[](2)(3)
            \Edge[Direct,bend=30,label=$w_{45}$](5)(4)
            \Edge[Direct,bend=30,label=$w_{54}$](4)(5)
        \end{tikzpicture}}}
    \qquad
    \scalebox{0.90}{{\begin{tikzpicture}
            \centering
            \Text[x=0,y=1.3,fontsize=\small]{$\mathcal{G}_m$};
            \Vertex[x=1,y=1,label=$1$,color=green,opacity=0.1,size=.5]{1}
            \Vertex[x=-1,y=1,label=$2$,color=green,opacity=0.1,size=.5]{2}
            \Vertex[x=-1,y=-1,label=$3$,color=green,opacity=0.1,size=.5]{3}
            \Vertex[x=0,y=-1,label=$4$,color=green,opacity=0.1,size=.5]{4}
            \Vertex[x=1,y=0,label=$5$,color=green,opacity=0.1,size=.5]{5}
            \Edge[](1)(2)
            \Edge[](2)(3)
            \Edge[Direct,bend=30,label=$w_{45}$](5)(4)
            \Edge[Direct,bend=30,label=$w_{54}$](4)(5)
        \end{tikzpicture}}}
    \qquad
    \scalebox{0.90}{{\begin{tikzpicture}
            \centering
            \Text[x=0,y=1.3,fontsize=\small]{$\mathcal{G}_0$};
            \Vertex[x=-.375,y=.375,label=$0$,color=red,opacity=0.1,size=.5]{L}
            \Vertex[x=1,y=1,label=$1$,color=green,opacity=0.1,size=.5]{1}
            \Vertex[x=0,y=-1,label=$4$,color=green,opacity=0.1,size=.5]{4}
            \Vertex[x=1,y=0,label=$5$,color=green,opacity=0.1,size=.5]{5}
            \Edge[Direct,color=red,label=$w_{10}$](L)(1)
            \Edge[Direct,color=red,label=$w_{40}$](L)(4)
            \Edge[Direct,color=red,label=$w_{50}$](L)(5)
        \end{tikzpicture}}}
    \caption{Example of a graph $\mathcal{G}$ with $m=5$, its
    sensing-to-sensing subgraph $\mathcal{G}_m$, and its
    source-to-sensing subgraph $\mathcal{G}_0$, showing the graph
    decomposition and assignment of $w_{ij}$. Each line in
    $\mathcal{G}_m$ represents two oppositely directed edges, as explicitly
    illustrated between nodes $4$ and $5$. The corresponding reciprocal
    weights need not be equal.}
    \label{Fig:network}
\end{figure}
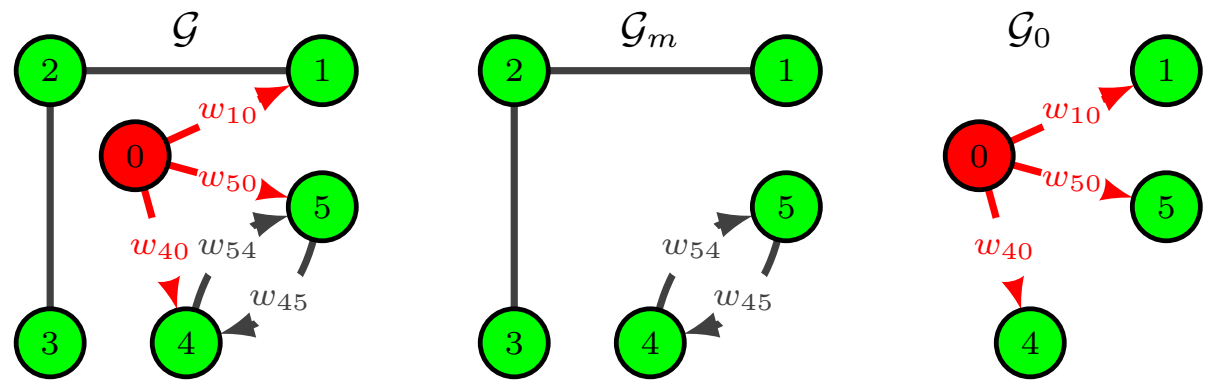

For analysis, we separate $\mathcal{G}$ into two subgraphs:
(i) the sensing subgraph
$\mathcal{G}_m=(\mathcal{V}_m,\mathcal{E}_m,\mathcal{W}_m)$ with
$\mathcal{V}_m=\{1,\dots,m\}$;
(ii) the source--sensing subgraph
$\mathcal{G}_0=(\mathcal{V}_0,\mathcal{E}_0,\mathcal{W}_0)$ with
$\mathcal{V}_0=\{0\}\cup\{\,i\in\mathcal{V}_m\mid(i,0)\in\mathcal{E}\,\}$.
The sensing subgraph $\mathcal{G}_m$ is assumed bidirected, i.e.,
$(i,j)\in\mathcal{E}_m$ if and only if $(j,i)\in\mathcal{E}_m$, although
the reciprocal weights $w_{ij}$ and $w_{ji}$ need not be equal. The
overall graph $\mathcal{G}$ remains directed through the source--sensing
links in $\mathcal{G}_0$; see Fig.~\ref{Fig:network}.
This decomposition simplifies the subsequent network-level analysis.

For the sensing subgraph $\mathcal{G}_m$, define the weighted in-degree
of node $i$ as
$d_i\coloneqq\sum_{j:(i,j)\in\mathcal{E}_m}w_{ij}$,
$i\in\mathcal{V}_m$, and let
$\mathbb{D}_m\coloneqq\diag\{d_1,\dots,d_m\}$. The corresponding
adjacency matrix $\mathbb{A}_m$ has entries
$[\mathbb{A}_m]_{ij}=w_{ij}$ whenever
$(i,j)\in\mathcal{E}_m$, and $[\mathbb{A}_m]_{ij}=0$ otherwise. The
Laplacian matrix associated with $\mathcal{G}_m$ is therefore given by
$\mathbb{L}_m\coloneqq\mathbb{D}_m-\mathbb{A}_m$.

The source--sensing connections in $\mathcal{G}_0$ are represented by
the diagonal matrix
$\mathbb{A}_0\coloneqq\diag\{w_{10},\dots,w_{m0}\}$, where
$[\mathbb{A}_0]_{ii}=w_{i0}$ if $(i,0)\in\mathcal{E}_0$ and
$[\mathbb{A}_0]_{ii}=0$ otherwise.

Let $\mathbb{W}\coloneqq\diag\{w_1,\dots,w_m\}$, where
$w_i\coloneqq d_i+w_{i0}$. The network coupling matrix is defined as
\begin{equation*}
    \mathbb{L}
    \coloneqq
    \mathbb{L}_m+\mathbb{A}_0
    =
    \mathbb{W}-\mathbb{A}_m.
\end{equation*}
We focus on the \emph{balanced} setting $\mathbb{W}=I_m$, under which
\begin{equation}\label{eq:ComNet:balance}
    (\mathbb{L}-\mathbb{A}_0)\mathbf{1}_m
    =
    \mathbf{0}_m
    \quad\Leftrightarrow\quad
    (\mathbb{A}_m+\mathbb{A}_0)\mathbf{1}_m
    =
    \mathbf{1}_m.
\end{equation}
Hence, the total incoming weight assigned by each sensing node to its
neighbors and the source is equal to one.\footnote{If
$\mathbb{W}\neq I_m$, the weights can be rescaled according to
$\tilde w_{ij}\coloneqq w_{ij}/w_i$ for
$(i,j)\in\mathcal{E}_m$ and
$\tilde w_{i0}\coloneqq w_{i0}/w_i$. This rescaling yields
$\tilde{\mathbb{W}}=I_m$ and preserves
\eqref{eq:ComNet:balance}, with
$\tilde{\mathbb{L}}=\tilde{\mathbb{L}}_m+\tilde{\mathbb{A}}_0$.
Although this normalization is not necessary for implementation, it
simplifies the stability analysis.}

\begin{remark}[Source reachability]\label{rem:reachability}
Suppose that at least one sensing node has direct access to the source,
that is, $w_{i0}>0$ for some $i\in\mathcal{V}_m$, and that every sensing
node can be reached from the source through a directed path. Under this
connectivity condition, the network coupling matrix
$\mathbb{L}=\mathbb{L}_m+\mathbb{A}_0$ is \emph{positive stable}; namely,
all its eigenvalues satisfy
$\Re\{\lambda(\mathbb{L})\}>0$.
\end{remark}

\section{Problem Formulation}\label{sec:ProblemFormulation}

Consider a distributed sensing network consisting of an \emph{unknown}
nonlinear source, indexed by $0$, and $m$ sensing nodes indexed by
$i\in\{1,\dots,m\}$. Each sensing node processes locally available
information and implements an adaptive observer to reconstruct the
source state while updating local estimates of its unknown dynamics.

The source dynamics are described by
\begin{equation}\label{eq:Problem:source}
    \dot{x}_0
    =
    A_0x_0
    +
    f_0(x_0)
    +
    B_0u_0
    +
    \delta_0,
\end{equation}
where $x_0\in\mathbb{R}^n$ denotes the source state and
$u_0\in\mathbb{R}^p$ is the \emph{unknown} source input. The \emph{unknown}
matrix $A_0\in\mathbb{R}^{n\times n}$ is assumed to be \emph{Hurwitz},
whereas $B_0\in\mathbb{R}^{n\times p}$ is \emph{known}. In particular, $A_0$
admits the decomposition
$A_0=A_\ast+\Delta A$, where $A_\ast$ is known and the uncertainty
$\Delta A$ satisfies $\|\Delta A\|_2\le a$ for some known constant
$a>0$.

The nonlinear term is parameterized as
$f_0(x_0)\coloneqq C_0g(x_0)$, where
$C_0\in\mathbb{R}^{n\times r}$ is an \emph{unknown} constant matrix and
$g:\mathbb{R}^n\to\mathbb{R}^r$ is a known smooth nonlinear mapping.
The disturbance $\delta_0\in\mathbb{R}^n$ is assumed to be bounded such
that $\|\delta_0(t)\|\le\delta^\ast$ for all $t\ge0$ and some constant
$\delta^\ast>0$.

\begin{assumption}\label{assmpt:Lipschitz}
The mapping $g:\mathbb{R}^n\to\mathbb{R}^r$ is Lipschitz continuous.
That is, there exists a constant $L_g>0$ such that
\begin{equation}\label{eq:g_Lipschitz}
    \|g(x)-g(y)\|
    \le
    L_g\|x-y\|,
    \qquad
    \forall x,y\in\mathbb{R}^n.
\end{equation}
\end{assumption}

\begin{assumption}\label{assmpt:source_boundedness}
The source input $u_0$ is continuously differentiable, and the source
state and input are bounded. That is, there exist constants
$x^\ast,u^\ast,\dot u^\ast>0$ such that
\begin{equation}\label{eq:source_bounds}
    \|x_0(t)\|
    \le
    x^\ast,
    \qquad
    \|u_0(t)\|
    \le
    u^\ast,
    \qquad
    \|\dot u_0(t)\|
    \le
    \dot u^\ast.
\end{equation}
for all $t\ge0$
\end{assumption}

The matrices $A_0$ and $C_0$, as well as the source input $u_0$, are
unknown to every sensing node. The objective is to reconstruct the
source state and adapt local estimates of its unknown dynamics using
only locally accessible information and communication with neighboring
nodes.

Each sensing node $i\in\{1,\dots,m\}$ implements
\begin{subequations}\label{eq:Problem:sensor_system}
\begin{align}
    \dot{x}_i
    &=
    H_ix_i
    +
    (\hat A_i-H_i)z_i
    +
    \hat C_i g(z_i)
    +
    B_i u_i,
    \label{eq:Problem:sensor}\\
    z_i
    &=
    \sum\nolimits_{j\in\mathcal{N}_i}
    w_{ij}x_j,
    \qquad
    \mathcal{N}_i\subseteq\mathcal{V},
    \label{eq:Problem:z}
\end{align}
\end{subequations}
where $x_i\in\mathbb{R}^n$ is node $i$'s estimate of the source state,
and $z_i\in\mathbb{R}^n$ is the weighted combination of the states
available from its in-neighbors.\footnote{In
\eqref{eq:Problem:sensor}, the nonlinear mapping $g(\cdot)$ is evaluated
at the locally available $z_i$ rather than the unknown source state
$x_0$.} The matrices
$\hat A_i\in\mathbb{R}^{n\times n}$ and
$\hat C_i\in\mathbb{R}^{n\times r}$ denote the local estimates of the
unknown matrices $A_0$ and $C_0$, respectively. All sensing nodes employ
the common known input matrix $B_i=B_0$, while
$H_i\in\mathbb{R}^{n\times n}$ is a user-selected Hurwitz matrix.

The local input of sensing node $i$ is adjusted according to
\begin{equation}\label{eq:Problem:u_update}
    \dot{u}_i
    \coloneqq
    \pi_{u,i}(\cdot)
    =
    -
    \gamma_{u_i}B_i^\top
    \left(
        P_ie_i
        +
        \sum\nolimits_{j\in\mathcal{N}_i\cap\mathcal V_m}
        [\mathbb{L}]_{ji}P_je_j
    \right),
\end{equation}
where $\gamma_{u_i}>0$ is the input adaptation gain,
$P_i=p_iP_n\succ0$ is the local weighting matrix determined by the
structured Lyapunov LMI in Lemma~\ref{lem:NL:LMI_H}, and
$e_i\coloneqq x_i-z_i$ denotes the local estimation error.

The update in \eqref{eq:Problem:u_update} is implemented without access
to the source input $u_0$ and without exchanging the neighboring inputs
$u_j$. The bounded rate $\dot u_0$ enters the input-mismatch dynamics as
an exogenous perturbation and will be accounted for in the subsequent
robustness analysis.

\begin{remark}[Locally available information]\label{rem:local}
    Each sensing node $i$ implements
    \eqref{eq:Problem:sensor}--\eqref{eq:Problem:z} and the adaptive update
    \eqref{eq:Problem:u_update} using only local and neighboring information.
    In particular, node $i$ receives $x_j$ to construct $z_i$ and $e_j$ to
    update $u_i$, without requiring the source input $u_0$ or the
    neighboring inputs $u_j$.
\end{remark}

\textbf{Main objective}: Develop distributed adaptive laws for
$\hat A_i$, $\hat C_i$, and the local input $u_i$ such that, under the
\emph{balanced} condition \eqref{eq:ComNet:balance} and the source
reachability condition in Remark~\ref{rem:reachability}, both the local
estimation error
$e_i\coloneqq x_i-z_i$ and the source-tracking error
$\epsilon_i\coloneqq x_i-x_0$ are ISS with respect to the predictor
mismatch and the bounded disturbance in \eqref{eq:Problem:source}. In
particular, the ISS property yields the asymptotic bounds
\begin{align}\label{eq:ProbFor:cont:ISS}
    \begin{aligned}
    \limsup_{t\to\infty}\|e_i(t)\|
    &\le
    \vartheta_{e,\eta}\eta^\ast
    +
    \vartheta_{e,\delta}\delta^\ast,\\
    \limsup_{t\to\infty}\|\epsilon_i(t)\|
    &\le
    \vartheta_{\epsilon,\eta}\eta^\ast
    +
    \vartheta_{\epsilon,\delta}\delta^\ast,
    \qquad i\in\{1,\dots,m\},
    \end{aligned}
\end{align}
for some constants
$\vartheta_{e,\eta},\vartheta_{e,\delta},
\vartheta_{\epsilon,\eta},\vartheta_{\epsilon,\delta}>0$.
Moreover, in the nominal case
$\delta_0(t)\equiv0$ and $\dot u_0(t)\equiv0$, both errors converge to
zero:
\begin{equation*}
    \lim_{t\to\infty}\|e_i(t)\|
    =
    \lim_{t\to\infty}\|\epsilon_i(t)\|
    =
    0,
    \qquad
    \forall i\in\{1,\dots,m\}.
\end{equation*}

To simplify the subsequent developments, we introduce several compact
notations. For any source signal $\beta_0\in\mathbb{R}^q$ and sensing-node
signals $\beta_i\in\mathbb{R}^q$, define
\begin{subequations}
\begin{align}
    \Bar{\beta}_0
    &\coloneqq
    \mathbf{1}_m\otimes\beta_0
    \in\mathbb{R}^{mq},
    \label{eq:ProbFor:beta0}\\
    \Bar{\beta}
    &\coloneqq
    [\beta_1^\top,\dots,\beta_m^\top]^\top
    \in\mathbb{R}^{mq}.
    \label{eq:ProbFor:betai}
\end{align}
\end{subequations}
Similarly, for matrices
$E_0,E_i,\hat E_i\in\mathbb{R}^{q_1\times q_2}$, let
\begin{subequations}
\begin{align}
    \mathbf{E}_0
    &\coloneqq
    I_m\otimes E_0
    \in\mathbb{R}^{mq_1\times mq_2},
    \label{eq:ProbFor:mat0}\\
    \mathbf{E}
    &\coloneqq
    \diag\{E_1,\dots,E_m\}
    \in\mathbb{R}^{mq_1\times mq_2},
    \label{eq:ProbFor:mati}\\
    \hat{\mathbf{E}}
    &\coloneqq
    \diag\{\hat E_1,\dots,\hat E_m\}
    \in\mathbb{R}^{mq_1\times mq_2}.
    \label{eq:ProbFor:matihat}
\end{align}
\end{subequations}
These definitions will be used to express the network dynamics and the
distributed adaptive laws in compact form throughout the continuous-time
stability analysis.

\section{Distributed Adaptive Identification}\label{sec:AdaptiveId}

This section presents the distributed adaptive identification framework.
We first formulate the network-level error dynamics and then derive
sufficient conditions for the Hurwitz stability of the resulting
coupling operator. Finally, we introduce the distributed adaptive laws
and examine their stability through Lyapunov analysis.

\subsection{Error Dynamics in Networked Systems}
\label{subsec:AdpId:Error}

Using the compact notation introduced in the preceding section, the
source dynamics can be written as
\begin{equation}\label{eq:AdpId:source}
    \dot{\Bar{x}}_0
    =
    \mathbf{A}_0\Bar{x}_0
    +
    \mathbf{C}_0\mathbf{g}(\Bar{x}_0)
    +
    \mathbf{B}_0\Bar{u}_0
    +
    \Bar{\delta}_0.
\end{equation}
The collective sensing-node dynamics are expressed as
\begin{subequations}\label{eq:AdpId:sensor_network}
\begin{align}
    \dot{\Bar{x}}
    &=
    \mathbf{H}\Bar{x}
    +
    (\hat{\mathbf{A}}-\mathbf{H})\Bar{z}
    +
    \hat{\mathbf{C}}\mathbf{g}(\Bar{z})
    +
    \mathbf{B}\Bar{u},
    \label{eq:AdpId:sensor}\\
    \Bar{z}
    &=
    (\mathbb{A}_m\otimes I_n)\Bar{x}
    +
    (\mathbb{A}_0\otimes I_n)\Bar{x}_0,
    \label{eq:AdpId:z}\\
    \dot{\Bar{u}}
    &=
    \Pi_u(\cdot).
    \label{eq:AdpId:u}
\end{align}
\end{subequations}
Here, $\Bar{x}_0$, $\Bar{u}_0$, and $\Bar{\delta}_0$ are defined
according to \eqref{eq:ProbFor:beta0}, whereas $\Bar{x}$, $\Bar{z}$,
and $\Bar{u}$ follow the stacking convention in
\eqref{eq:ProbFor:betai}. 
The matrices $\mathbf{A}_0$, $\mathbf{B}_0$, and $\mathbf{C}_0$ are
constructed as in \eqref{eq:ProbFor:mat0}, while $\mathbf{H}$ follows
\eqref{eq:ProbFor:mati}.
Similarly, the adaptive estimates
$\hat{\mathbf{A}}$ and $\hat{\mathbf{C}}$ are defined according to
\eqref{eq:ProbFor:matihat}. Since all sensing nodes use the common input
matrix $B_i=B_0$, it follows that $\mathbf{B}\equiv\mathbf{B}_0$.

The stacked nonlinear and input-update terms are given by
\begin{equation*}
    \begin{aligned}
    \mathbf{g}(\Bar{x}_0)
    &\coloneqq
    \mathbf{1}_m\otimes g(x_0),
    \\
    \mathbf{g}(\Bar{z})
    &\coloneqq
    [g(z_1)^\top,\dots,g(z_m)^\top]^\top,
    \\
    \Pi_u(\cdot)
    &\coloneqq
    [\pi_{u,1}(\cdot)^\top,\dots,
    \pi_{u,m}(\cdot)^\top]^\top.
    \end{aligned}
\end{equation*}
The matrices $\mathbb{A}_m$ and $\mathbb{A}_0$ appearing in
\eqref{eq:AdpId:z} are the network weighting matrices introduced in
Section~\ref{sec:ComNetwork}.

For notational convenience, define the ideal parameter estimates as
\begin{equation}\label{eq:AdpId:ideal_param}
    \hat{\mathbf{A}}^\ast
    \coloneqq
    \mathbf{A}_0,
    \qquad
    \hat{\mathbf{C}}^\ast
    \coloneqq
    \mathbf{C}_0.
\end{equation}
The stacked source-tracking error is defined by
\begin{equation}\label{eq:AdpId:tracking_error}
    \Bar{\epsilon}
    \coloneqq
    \Bar{x}-\Bar{x}_0.
\end{equation}
Under the balanced condition \eqref{eq:ComNet:balance}, the stacked local
estimation error satisfies
\begin{subequations}
\begin{align}
    \Bar e
    \coloneqq
    \Bar{x}-\Bar{z}
    &=
    (\mathbb L\otimes I_n)\Bar x
    -
    (\mathbb A_0\otimes I_n)\Bar x_0,
    \label{eq:AdpId:error}\\
    &=
    (\mathbb L\otimes I_n)
    (\Bar x-\Bar x_0),
    \nonumber\\
    &=
    (\mathbb L\otimes I_n)\Bar\epsilon.
    \label{eq:AdpId:error_tracking_relation}
\end{align}
\end{subequations}
Here, we used
$(\mathbb L\otimes I_n)\Bar x_0
=(\mathbb A_0\otimes I_n)\Bar x_0$.
Because $\mathbb L$ is nonsingular under the source reachability
condition in Remark~\ref{rem:reachability}, it follows that
\begin{equation}\label{eq:AdpId:tracking_from_error}
    \Bar\epsilon
    =
    (\mathbb L^{-1}\otimes I_n)\Bar e.
\end{equation}
Furthermore, using \eqref{eq:AdpId:z} together with
$(\mathbb A_m+\mathbb A_0)\mathbf 1_m=\mathbf 1_m$, we obtain
\begin{equation}\label{eq:AdpId:z_tracking_relation}
\begin{aligned}
    \Bar z-\Bar x_0
    &=
    (\mathbb A_m\otimes I_n)\Bar x
    +
    (\mathbb A_0\otimes I_n)\Bar x_0
    -
    \Bar x_0\\
    &=
    (\mathbb A_m\otimes I_n)
    (\Bar x-\Bar x_0)\\
    &=
    (\mathbb A_m\mathbb L^{-1}\otimes I_n)\Bar e.
\end{aligned}
\end{equation}
Next, define the parameter estimation errors, input mismatch, and
nonlinear mapping mismatch as
\begin{equation}\label{eq:AdpId:par_error}
\begin{aligned}
    \Phi
    &\coloneqq
    \hat{\mathbf A}-\mathbf A_0,
    &
    \Psi
    &\coloneqq
    \hat{\mathbf C}-\mathbf C_0,\\
    \Bar r
    &\coloneqq
    \Bar u-\Bar u_0,
    &
    \Delta\mathbf g
    &\coloneqq
    \mathbf g(\Bar z)-\mathbf g(\Bar x_0).
\end{aligned}
\end{equation}
Also, let
$\Omega\coloneqq\mathbb L\otimes I_n$.
Differentiating \eqref{eq:AdpId:error} along the source and sensing-node
dynamics gives
\begin{align}\label{eq:AdpId:error_step1}
\begin{aligned}
    \dot{\Bar e}
    &=
    \Omega\dot{\Bar x}
    -
    (\mathbb A_0\otimes I_n)\dot{\Bar x}_0\\
    &=
    \Omega
    \Bigl[
        \mathbf H\Bar x
        +
        (\hat{\mathbf A}-\mathbf H)\Bar z
        +
        \hat{\mathbf C}\mathbf g(\Bar z)
        +
        \mathbf B_0\Bar u
    \Bigr]\\
    &\quad
    -
    (\mathbb A_0\otimes I_n)
    \Bigl[
        \mathbf A_0\Bar x_0
        +
        \mathbf C_0\mathbf g(\Bar x_0)
        +
        \mathbf B_0\Bar u_0
        +
        \Bar\delta_0
    \Bigr].
\end{aligned}
\end{align}
For any replicated source quantity
$\Bar\beta_0=\mathbf 1_m\otimes\beta_0$, the balanced condition implies
\begin{equation*}
    (\mathbb A_0\otimes I_q)\Bar\beta_0
    =
    (\mathbb L\otimes I_q)\Bar\beta_0.
\end{equation*}
Consequently, \eqref{eq:AdpId:error_step1} can be rewritten as
\begin{align}\label{eq:AdpId:error_step2}
\begin{aligned}
    \dot{\Bar e}
    &=
    \Omega
    \Bigl[
        \mathbf H(\Bar x-\Bar z)
        +
        \hat{\mathbf A}\Bar z
        -
        \mathbf A_0\Bar x_0
    \Bigr]
    +
    \Omega
    \Bigl[
        \hat{\mathbf C}\mathbf g(\Bar z)
        -
        \mathbf C_0\mathbf g(\Bar x_0)
    \Bigr]\\
    &\quad
    +
    \Omega\mathbf B_0
    (\Bar u-\Bar u_0)
    -
    \Omega\Bar\delta_0.
\end{aligned}
\end{align}
Using
$\Delta\mathbf g
=\mathbf g(\Bar z)-\mathbf g(\Bar x_0)$,
$\hat{\mathbf A}=\mathbf A_0+\Phi$, and
$\hat{\mathbf C}=\mathbf C_0+\Psi$, the linear and nonlinear parameter
terms admit the decompositions
\begin{align}\label{eq:AdpId:AC_decomposition}
\begin{aligned}
    \hat{\mathbf A}\Bar z
    -
    \mathbf A_0\Bar x_0
    &=
    \mathbf A_0(\Bar z-\Bar x_0)
    +
    \Phi\Bar z,\\
    \hat{\mathbf C}\mathbf g(\Bar z)
    -
    \mathbf C_0\mathbf g(\Bar x_0)
    &=
    \Psi\mathbf g(\Bar z)
    +
    \mathbf C_0\Delta\mathbf g.
\end{aligned}
\end{align}
Substituting \eqref{eq:AdpId:AC_decomposition} into
\eqref{eq:AdpId:error_step2} and using
$\Bar e=\Bar x-\Bar z$ gives
\begin{align}\label{eq:AdpId:error_step3}
\begin{aligned}
    \dot{\Bar e}
    &=
    \Omega\mathbf H\Bar e
    +
    \Omega\mathbf A_0
    (\Bar z-\Bar x_0)
    +
    \Omega
    \Bigl[
        \Phi\Bar z
        +
        \Psi\mathbf g(\Bar z)
        +
        \mathbf B_0\Bar r
    \Bigr]\\
    &\quad
    +
    \Omega\mathbf C_0\Delta\mathbf g
    -
    \Omega\Bar\delta_0.
\end{aligned}
\end{align}
From \eqref{eq:AdpId:z_tracking_relation} and the mixed-product property
$(A\otimes B)(C\otimes D)=(AC)\otimes(BD)$, it follows that
\begin{align*}
    \Omega\mathbf A_0
    (\Bar z-\Bar x_0)
    &=
    (\mathbb L\otimes I_n)
    (I_m\otimes A_0)
    (\mathbb A_m\mathbb L^{-1}\otimes I_n)
    \Bar e\\
    &=
    (\mathbb L\otimes A_0)
    (\mathbb A_m\mathbb L^{-1}\otimes I_n)
    \Bar e\\
    &=
    (\mathbb L\mathbb A_m\mathbb L^{-1}\otimes A_0)
    \Bar e.
\end{align*}
Under the balanced condition \eqref{eq:ComNet:balance},
$\mathbb A_m=I_m-\mathbb L$. Therefore,
\begin{equation*}
    \mathbb L\mathbb A_m
    =
    \mathbb L(I_m-\mathbb L)
    =
    (I_m-\mathbb L)\mathbb L
    =
    \mathbb A_m\mathbb L,
\end{equation*}
which implies
$\mathbb L\mathbb A_m\mathbb L^{-1}=\mathbb A_m$. Consequently,
\begin{align}\label{eq:AdpId:A_error_relation}
\begin{aligned}
    \Omega\mathbf A_0
    (\Bar z-\Bar x_0)
    &=
    (\mathbb A_m\otimes A_0)\Bar e\\
    &=
    (\mathbb A_m\otimes I_n)
    (I_m\otimes A_0)\Bar e\\
    &=
    (\mathbb A_m\otimes I_n)
    \mathbf A_0\Bar e.
\end{aligned}
\end{align}
The complete continuous-time estimation-error dynamics can thus be
written as
\begin{equation}\label{eq:AdpId:errordyn}
    \dot{\Bar e}
    =
    \mathcal H\Bar e
    +
    \Omega\Bar\eta
    +
    \Omega\mathbf C_0\Delta\mathbf g
    -
    \Omega\Bar\delta_0,
\end{equation}
where
\begin{subequations}
\begin{align}
    \mathcal H
    &\coloneqq
    \Omega\mathbf H
    +
    (\mathbb A_m\otimes I_n)\mathbf A_0,
    \label{eq:AdpId:H}\\
    \Bar\eta
    &\coloneqq
    \Phi\Bar z
    +
    \Psi\mathbf g(\Bar z)
    +
    \mathbf B_0\Bar r.
    \label{eq:AdpId:eta}
\end{align}
\end{subequations}
Equivalently, the adaptive predictor mismatch can be expressed as
$\Bar\eta=\Xi\Bar\xi$, where
\begin{equation*}
    \Xi
    \coloneqq
    [\,\Phi,\Psi,\mathbf B_0\,],
    \qquad
    \Bar\xi
    \coloneqq
    [\,\Bar z^\top,\mathbf g(\Bar z)^\top,\Bar r^\top\,]^\top.
\end{equation*}

The error dynamics in \eqref{eq:AdpId:errordyn} contain the nonlinear
mismatch $\Delta\mathbf g$, which depends on the difference between the
locally available aggregate $\Bar z$ and the source state $\Bar x_0$.
To express this term directly in terms of the estimation error $\Bar e$,
we invoke Assumption~\ref{assmpt:Lipschitz} and
\eqref{eq:AdpId:z_tracking_relation}. Accordingly,
\begin{align}
    \|\Delta\mathbf g\|^2
    =
    \sum_{i=1}^m
    \|g(z_i)-g(x_0)\|^2
    &\le
    L_g^2
    \sum_{i=1}^m
    \|z_i-x_0\|^2
    \nonumber\\
    &=
    L_g^2
    \|\Bar z-\Bar x_0\|^2
    \le
    L_g^2
    \|\mathbb A_m\mathbb L^{-1}\otimes I_n\|_2^2
    \|\Bar e\|^2.
    \label{eq:AdpId:g_mismatch_bound}
\end{align}
Thus, the nonlinear mismatch is bounded by the estimation error and can
be incorporated into the subsequent stability analysis. 
We first examine the properties of the coupling operator $\mathcal H$
defined in \eqref{eq:AdpId:H}.

\subsection{Stability of the Coupling Operator $\mathcal{H}$}
\label{subsec:AdpId:H}

Before introducing the adaptive laws, we establish sufficient conditions
under which the coupling operator $\mathcal H$ is Hurwitz. This property
ensures exponential stability of the nominal error dynamics
$\dot{\Bar e}=\mathcal H\Bar e$ and provides the foundation for the
subsequent adaptive stability analysis.

For a square matrix $M$, define its Euclidean logarithmic norm as
\begin{equation*}
    \mu_2(M)
    \coloneqq
    \lambda_{\max}
    \left(
        \frac{M+M^\top}{2}
    \right).
\end{equation*}

\begin{proposition}\label{prop:NL:H_hurwitz}
Let $\mathcal H$ be defined by \eqref{eq:AdpId:H}. If
\begin{equation}\label{eq:NL:H_hurwitz:cond}
    \mu_2(\Omega\mathbf H)
    +
    \|\mathbb A_m\|_2\|A_0\|_2
    <
    0,
\end{equation}
then $\mathcal H$ is Hurwitz.
\end{proposition}

Condition~\eqref{eq:NL:H_hurwitz:cond} reveals how the communication
network, the local observer dynamics, and the source dynamics jointly
affect stability. In particular, the dissipativity of
$\Omega\mathbf H$ must dominate the contribution of the source dynamics,
which is scaled by the network gain $\|\mathbb A_m\|_2$. Hence, stronger
dissipation in the local observer dynamics can accommodate a larger
induced norm $\|A_0\|_2$ while preserving the Hurwitz stability of
$\mathcal H$.

Although Proposition~\ref{prop:NL:H_hurwitz} provides a direct and
easily verifiable condition, the use of induced norms may introduce
conservatism. The following structured Lyapunov condition retains the
full matrix structure of the coupling operator and can therefore provide
a less conservative stability certificate.

\begin{lemma}[Structured Lyapunov LMI]\label{lem:NL:LMI_H}
Let $\mathcal H$ be defined by \eqref{eq:AdpId:H}, and fix
$P_n=P_n^\top\succ0$. If there exists a diagonal matrix
$P_m=\diag\{p_1,\dots,p_m\}\succ0$ such that, with
$P_{mn}\coloneqq P_m\otimes P_n=\diag\{P_1,\dots,P_m\}$ and
$P_i\coloneqq p_iP_n\succ0$, the structured Lyapunov LMI
\begin{equation}\label{eq:NL:structured_LMI}
    \mathcal H^\top P_{mn}
    +
    P_{mn}\mathcal H
    \prec0
\end{equation}
holds, then $\mathcal H$ is Hurwitz. Moreover, the quadratic function
\begin{equation*}
    V(\Bar e)
    =
    \Bar e^\top P_{mn}\Bar e
\end{equation*}
satisfies $\dot V(\Bar e)<0$ for every $\Bar e\neq0$ along the nominal
error dynamics
$\dot{\Bar e}=\mathcal H\Bar e$.
\end{lemma}

We next account for the norm-bounded uncertainty in $A_0$ and derive a
robust structured LMI that guarantees Hurwitz stability for every
admissible realization of the source dynamics.

\begin{corollary}[Robust structured LMI]\label{cor:NL:robust_H}
Let $A_0=A_\ast+\Delta A$, where $\|\Delta A\|_2\le a$, and define
\begin{equation*}
    \mathcal H_\ast
    \coloneqq
    \Omega\mathbf H
    +
    (\mathbb A_m\otimes I_n)\mathbf A_\ast,
    \qquad
    \Delta\mathcal H
    \coloneqq
    (\mathbb A_m\otimes I_n)\Delta\mathbf A,
\end{equation*}
where
$\mathbf A_\ast=I_m\otimes A_\ast$ and
$\Delta\mathbf A=I_m\otimes\Delta A$. Thus,
$\mathcal H=\mathcal H_\ast+\Delta\mathcal H$.

Fix $P_n=P_n^\top\succ0$ and a scalar $\varepsilon>0$. If there exists
a diagonal matrix $P_m=\diag\{p_1,\dots,p_m\}\succ0$ such that, with
$P_{mn}\coloneqq P_m\otimes P_n$,
\begin{equation}\label{eq:NL:robust_LMI}
    \begin{bmatrix}
        \mathcal H_\ast^\top P_{mn}
        +
        P_{mn}\mathcal H_\ast
        +
        \dfrac{a^2\|\mathbb A_m\|_2^2}{\varepsilon}I_{mn}
        &
        P_{mn}
        \\
        P_{mn}
        &
        -\dfrac{1}{\varepsilon}I_{mn}
    \end{bmatrix}
    \prec0,
\end{equation}
then $\mathcal H$ is Hurwitz for every admissible uncertainty
$\Delta A$ satisfying $\|\Delta A\|_2\le a$.
\end{corollary}

The results above provide complementary conditions for verifying the
Hurwitz stability of $\mathcal H$ in the presence of network coupling
and model uncertainty. Proposition~\ref{prop:NL:H_hurwitz} offers a
direct norm-based test, Lemma~\ref{lem:NL:LMI_H} provides a structured
Lyapunov certificate, and Corollary~\ref{cor:NL:robust_H} extends this
certificate to uncertain source dynamics. With these conditions in
place, we next formulate the distributed adaptive laws and analyze the
stability of the complete error system.

\subsection{Closed-Loop Stability Under Adaptive Laws}
\label{subsec:AdpId:Stability}

Having established conditions under which $\mathcal H$ is Hurwitz, we
now analyze the complete adaptive network. We first introduce the
distributed adaptive laws and then establish a general Lyapunov bound,
followed by boundedness and asymptotic convergence for the unperturbed
closed-loop system.

Recall from \eqref{eq:AdpId:eta} that the adaptive predictor mismatch is
\begin{equation}\label{eq:AdpId:eta_recall}
    \Bar\eta
    =
    \Phi\Bar z
    +
    \Psi\mathbf g(\Bar z)
    +
    \mathbf B_0\Bar r.
\end{equation}
Define the network-weighted error direction as
\begin{equation}\label{eq:AdpId:zeta}
    \zeta
    \coloneqq
    \Omega^\top P_{mn}\Bar e
    \in\mathbb{R}^{mn},
    \qquad
    \Omega=\mathbb L\otimes I_n,
\end{equation}
where
$P_{mn}=P_m\otimes P_n=\diag\{P_1,\dots,P_m\}\succ0$
satisf Lemma~\ref{lem:NL:LMI_H}. Under the normalized-weight condition,
the $i$th block of \eqref{eq:AdpId:zeta} is
\begin{equation}\label{eq:AdpId:zeta_i}
    \zeta_i
    =
    P_ie_i
    +
    \sum\nolimits_{j\in\mathcal N_i\cap\mathcal V_m}
    [\mathbb L]_{ji}P_je_j.
\end{equation}
Thus, \(\zeta_i\) coincides with the locally available error direction
used in \eqref{eq:Problem:u_update}.

The adaptive laws implemented by sensing node $i$ are
\begin{subequations}\label{eq:AdpId:updates}
\begin{align}
    \dot{\hat A}_i
    &=
    -\gamma_{A_i}\zeta_i z_i^\top,
    &
    \dot\Phi_i&=\dot{\hat A}_i,
    \label{eq:AdpId:update_A}\\
    \dot{\hat C}_i
    &=
    -\gamma_{C_i}\zeta_i g(z_i)^\top,
    &
    \dot\Psi_i&=\dot{\hat C}_i,
    \label{eq:AdpId:update_C}\\
    \dot u_i
    &=
    -\gamma_{u_i}B_0^\top\zeta_i,
    \label{eq:AdpId:update_u}
\end{align}
\end{subequations}
where
$\gamma_{A_i},\gamma_{C_i},\gamma_{u_i}>0$ are scalar adaptation gains.
Unlike the discrete-time formulation, these update laws do not require
normalization.

The corresponding network-level quantities are
\begin{equation*}
\begin{aligned}
    \Phi
    &=
    \diag\{\Phi_1,\dots,\Phi_m\},
    &
    \Gamma_A
    &=
    \diag\{\gamma_{A_1},\dots,\gamma_{A_m}\}
    \otimes I_n,\\
    \Psi
    &=
    \diag\{\Psi_1,\dots,\Psi_m\},
    &
    \Gamma_C
    &=
    \diag\{\gamma_{C_1},\dots,\gamma_{C_m}\}
    \otimes I_n,\\
    \Bar r
    &=
    [r_1^\top,\dots,r_m^\top]^\top,
    &
    \Gamma_u
    &=
    \diag\{\gamma_{u_1},\dots,\gamma_{u_m}\}
    \otimes I_p.
\end{aligned}
\end{equation*}

We now state the principal stability result for the complete adaptive
error system.

\begin{theorem}\label{thm:AdpId:CL}
Suppose that the conditions of Lemma~\ref{lem:NL:LMI_H} hold, and define
\begin{equation*}
    Q_{mn}
    \coloneqq
    -\left(
        \mathcal H^\top P_{mn}
        +
        P_{mn}\mathcal H
    \right)
    \succ0.
\end{equation*}
Consider the error dynamics \eqref{eq:AdpId:errordyn} under the adaptive
laws \eqref{eq:AdpId:updates}. Suppose that
\begin{equation}\label{eq:AdpId:nonlinear_small_gain}
    \lambda_{\min}(Q_{mn})
    >
    2\|\Omega^\top P_{mn}\|_2
    \|C_0\|_2
    L_g
    \|\mathbb A_m\mathbb L^{-1}\otimes I_n\|_2.
\end{equation}
Then:
\begin{enumerate}[leftmargin=*]
    \item[i.] The composite Lyapunov function satisfies
    \begin{align}
        \dot V
        \le{}&
        -\alpha\|\Bar e\|^2
        +
        c_\delta\|\Bar\delta_0\|^2
        +
        \frac{\varepsilon_r}{\underline\gamma_u}
        \|\Bar r\|^2
        +
        c_u\|\dot{\Bar u}_0\|^2,
        \label{eq:AdpId:Vdot_main}
    \end{align}
    for some constants
    $\alpha,c_\delta,c_u,\varepsilon_r>0$.

    \item[ii.] In the nominal case
    $\Bar\delta_0(t)\equiv\mathbf0_{mn}$ and
    $\dot u_0(t)\equiv0$, the signals
    $\Bar e$, $\Phi$, $\Psi$, and $\Bar r$ remain bounded for all
    $t\ge0$, and
    \begin{equation}\label{eq:AdpId:error_convergence}
        \lim_{t\to\infty}\|\Bar e(t)\|
        =
        \lim_{t\to\infty}\|\Bar\epsilon(t)\|
        =
        0.
    \end{equation}

    \item[iii.] Under the conditions of item~\textnormal{(ii)}, the
    adaptive update rates satisfy
    \begin{equation}\label{eq:AdpId:update_convergence}
        \lim_{t\to\infty}\|\dot\Phi(t)\|_F
        =
        \lim_{t\to\infty}\|\dot\Psi(t)\|_F
        =
        \lim_{t\to\infty}\|\dot{\Bar r}(t)\|
        =
        0.
    \end{equation}
\end{enumerate}
\end{theorem}

\begin{remark}[Persistent excitation]\label{rem:PE}
Exact identification of $A_0$ and $C_0$ generally requires an
appropriate persistent excitation and identifiability condition on the
combined regressors $z_i$ and $g(z_i)$. Without such a condition,
Theorem~\ref{thm:AdpId:CL} guarantees bounded parameter estimates and
asymptotic state reconstruction, but not convergence of the parameter
estimates to their true values.
\end{remark}

\begin{corollary}[ISS robustness]\label{cor:AdpId:ISS}
Under the conditions of Theorem~\ref{thm:AdpId:CL}(i), suppose that
\begin{equation*}
    \|\Bar\eta(t)\|
    \le
    \eta^\ast,
    \qquad
    \|\Bar\delta_0(t)\|
    \le
    \delta^\ast,
    \qquad
    \forall t\ge0.
\end{equation*}
Then there exist positive constants
$\vartheta_{e,0}$, $\vartheta_{e,\eta}$,
$\vartheta_{e,\delta}$,
$\vartheta_{\epsilon,0}$,
$\vartheta_{\epsilon,\eta}$, and
$\vartheta_{\epsilon,\delta}$, and a constant $\lambda>0$, such that
\begin{subequations}\label{eq:AdpId:ISS_finite}
\begin{align}
    \|\Bar e(t)\|
    &\le
    \vartheta_{e,0}e^{-\lambda t/2}\|\Bar e(0)\|
    +
    \vartheta_{e,\eta}
    \|\Bar\eta\|_{[0,t],\infty}
    +
    \vartheta_{e,\delta}
    \|\Bar\delta_0\|_{[0,t],\infty},
    \label{eq:AdpId:ISS_finite_e}\\
    \|\Bar\epsilon(t)\|
    &\le
    \vartheta_{\epsilon,0}
    e^{-\lambda t/2}\|\Bar e(0)\|
    +
    \vartheta_{\epsilon,\eta}
    \|\Bar\eta\|_{[0,t],\infty} 
    +
    \vartheta_{\epsilon,\delta}
    \|\Bar\delta_0\|_{[0,t],\infty}.
    \label{eq:AdpId:ISS_finite_epsilon}
\end{align}
\end{subequations}
Consequently,
\begin{subequations}\label{eq:AdpId:ISS}
\begin{align}
    \limsup_{t\to\infty}\|\Bar e(t)\|
    &\le
    \vartheta_{e,\eta}\eta^\ast
    +
    \vartheta_{e,\delta}\delta^\ast,
    \label{eq:AdpId:ISS_e}\\
    \limsup_{t\to\infty}\|\Bar\epsilon(t)\|
    &\le
    \vartheta_{\epsilon,\eta}\eta^\ast
    +
    \vartheta_{\epsilon,\delta}\delta^\ast.
    \label{eq:AdpId:ISS_epsilon}
\end{align}
\end{subequations}
\end{corollary}

Theorem~\ref{thm:AdpId:CL} establishes boundedness of the adaptive
variables and asymptotic convergence of the estimation and tracking
errors in the absence of external perturbations. Exact parameter
identification requires the additional excitation condition described
in Remark~\ref{rem:PE}.

\section{Numerical Results}
\label{sec:NR}

\begin{figure}[h!]
    \centering
    \scalebox{0.90}{{\begin{tikzpicture}
            \centering
            \Text[x=0,y=1.5]{$\mathcal{G}_{\mathrm{s}}\coloneqq$ Star};
            \Vertex[x=0,y=0,label=$0$,color=red,opacity=0.1,size=.5]{L}
            \Vertex[x=1,y=1,label=$1$,color=green,opacity=0.1,size=.5]{1}
            \Vertex[x=-1,y=1,label=$2$,color=green,opacity=0.1,size=.5]{2}
            \Vertex[x=-1,y=-1,label=$3$,color=green,opacity=0.1,size=.5]{3}
            \Vertex[x=1,y=-1,label=$4$,color=green,opacity=0.1,size=.5]{4}
            \Edge[Direct,color=red,label=$1.0$](L)(1)
            \Edge[Direct,color=red,label=$1.0$](L)(2)
            \Edge[Direct,color=red,label=$1.0$](L)(3)
            \Edge[Direct,color=red,label=$1.0$](L)(4)
        \end{tikzpicture}}}
    \qquad
    \scalebox{0.90}{{\begin{tikzpicture}
            \centering
            \Text[x=0,y=1.5]{$\mathcal{G}_{\mathrm{c}}\coloneqq$ Cyclic};
            \Vertex[x=0,y=0,label=$0$,color=red,opacity=0.1,size=.5]{L}
            \Vertex[x=1,y=1,label=$1$,color=green,opacity=0.1,size=.5]{1}
            \Vertex[x=-1,y=1,label=$2$,color=green,opacity=0.1,size=.5]{2}
            \Vertex[x=-1,y=-1,label=$3$,color=green,opacity=0.1,size=.5]{3}
            \Vertex[x=1,y=-1,label=$4$,color=green,opacity=0.1,size=.5]{4}
            \Edge[Direct,color=red,label=$0.4$](L)(1)
            \Edge[Direct,color=red,label=$0.4$](L)(2)
            \Edge[Direct,color=red,label=$0.4$](L)(3)
            \Edge[Direct,color=red,label=$0.4$](L)(4)
            \draw[<->,ultra thick, latex' -latex'] (1) -- node[midway, fill=white, inner sep=2pt]{\small $0.3$} (2);
            \draw[<->,ultra thick, latex' -latex'] (2) -- node[midway, fill=white, inner sep=2pt]{\small $0.3$} (3);
            \draw[<->,ultra thick, latex' -latex'] (3) -- node[midway, fill=white, inner sep=2pt]{\small $0.3$} (4);
            \draw[<->,ultra thick, latex' -latex'] (4) -- node[midway, fill=white, inner sep=2pt]{\small $0.3$} (1);
        \end{tikzpicture}}}
    \qquad
    \scalebox{0.90}{{\begin{tikzpicture}
            \centering
            \Text[x=0,y=1.5]{$\mathcal{G}_{\mathrm{p}}\coloneqq$ Path};
            \Vertex[x=0,y=0,label=$0$,color=red,opacity=0.1,size=.5]{L}
            \Vertex[x=1,y=1,label=$1$,color=green,opacity=0.1,size=.5]{1}
            \Vertex[x=-1,y=1,label=$2$,color=green,opacity=0.1,size=.5]{2}
            \Vertex[x=-1,y=-1,label=$3$,color=green,opacity=0.1,size=.5]{3}
            \Vertex[x=1,y=-1,label=$4$,color=green,opacity=0.1,size=.5]{4}
            \Edge[Direct,color=red,label=$1.0$](L)(1)
            \draw[<->,ultra thick, -latex'] (1) -- node[midway, fill=white, inner sep=2pt]{\small $1.0$} (2);
            \draw[<->,ultra thick, -latex'] (2) -- node[midway, fill=white, inner sep=2pt]{\small $1.0$} (3);
            \draw[<->,ultra thick, -latex'] (3) -- node[midway, fill=white, inner sep=2pt]{\small $1.0$} (4);
        \end{tikzpicture}}}
    \caption{Three network topologies $(\mathcal{G}_{\mathrm{s}}, \mathcal{G}_{\mathrm{c}}, \mathcal{G}_{\mathrm{p}})$ with weights used in the simulations.}
    \label{Fig:topology}
\end{figure}
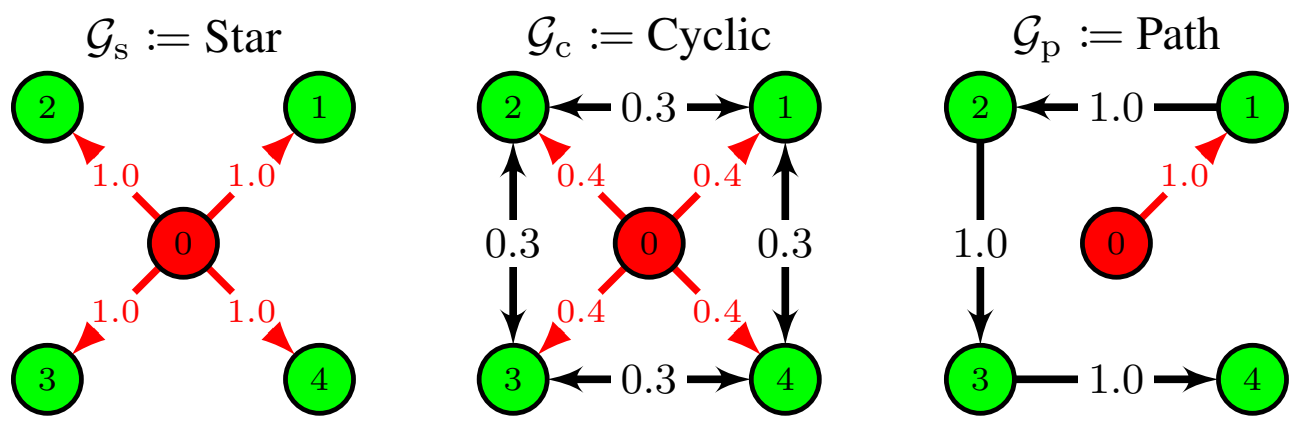

This section evaluates the proposed continuous-time distributed adaptive
identification method over three directed communication topologies:
star--$\mathcal{G}_{\mathrm{s}}$, cyclic--$\mathcal{G}_{\mathrm{c}}$,
and path--$\mathcal{G}_{\mathrm{p}}$. The corresponding graphs and
communication weights are shown in Fig.~\ref{Fig:topology}. All three
topologies satisfy the balanced condition
\eqref{eq:ComNet:balance} and the source reachability condition in
Remark~\ref{rem:reachability}. We first consider networks with $m=4$
sensing nodes under nominal and perturbed conditions. We then compare
the stability certificates developed for the coupling operator
$\mathcal H$ and investigate the computational scalability of the
method for networks containing up to $500$ sensing nodes.

\subsection{Simulation Setting}

The unknown nonlinear source evolves according to
\eqref{eq:Problem:source}, with
\begin{equation*}
    \small{
    A_0
    =
    \begin{bmatrix}
        -0.80 & 0.30\\
        -0.20 & -0.65
    \end{bmatrix},
    \;
    C_0
    =
    \begin{bmatrix}
        0.06 & 0\\
        0.10 & 0.05
    \end{bmatrix},
    \;
    B_0
    =
    \begin{bmatrix}
        0\\
        0.60
    \end{bmatrix}.}
\end{equation*}
The matrix $A_0$ is Hurwitz, with spectral abscissa
$\max\Re\{\lambda(A_0)\}=-0.725$. The matrices $A_0$ and $C_0$ are
unknown to every sensing node, whereas the input matrix $B_0$ is known.
The nonlinear mapping is selected as
\begin{equation*}
    g(x)
    =
    \begin{bmatrix}
        \sin(x^{(1)})\\
        \tanh(x^{(2)})
    \end{bmatrix}.
\end{equation*}
Since the derivatives of both scalar components are bounded in
magnitude by one, $g(\cdot)$ is globally Lipschitz with $L_g=1$.

Two simulation scenarios are considered.
\begin{enumerate}[leftmargin=*]
    \item The first is the
    \emph{nominal case}, which illustrates the asymptotic convergence result
    of Theorem~\ref{thm:AdpId:CL}. 
    The unknown source input is constant, with $u_0(t)=0.50$ and
    $\dot u_0(t)=0$, and the disturbance is absent, with
    $\delta_0(t)=\mathbf 0_2$.
    Thus, the nominal setting satisfies the unperturbed conditions of
    Theorem~\ref{thm:AdpId:CL} while retaining an unknown nonzero input that
    must be reproduced by the local adaptive inputs.

    \item The second scenario is the \emph{ISS case}, in which the unknown source
    input and its time derivative are given by
    \begin{align*}
        u_0(t) &= 0.60\sin(0.35t) + 0.25\cos(0.20t), \\
        \dot u_0(t) &= 0.21\cos(0.35t) - 0.05\sin(0.20t).
    \end{align*}
    Consequently, its rate of variation satisfies
    $\|\dot u_0(t)\| \le u^\ast \coloneqq 0.26$.

    The source is also affected by the bounded disturbance
    \begin{equation*}
        \delta_0(t)
        =
        0.01
        \begin{bmatrix}
            0.70\sin(0.45t)\\
            0.50\cos(0.30t)
        \end{bmatrix},
    \end{equation*}
    which satisfies
    \begin{equation*}
        \|\delta_0(t)\|
        \le
        0.01\sqrt{0.70^2+0.50^2}
        \approx
        8.60\times10^{-3}.
    \end{equation*}
    This scenario is used to examine the robustness of the estimation and
    tracking errors in the presence of a time-varying unknown input and an
    external disturbance.
\end{enumerate}

For both cases, the initial conditions are
$x_0(0)=[1,-0.5]^\top$ and
$x_i(0)\sim\mathcal N(0,I_2)$. The adaptive estimates are initialized
as
$\hat A_i(0)=0_{2\times2}$ and
$\hat C_i(0)=0_{2\times2}$, while the local adaptive inputs satisfy
$u_i(0)=0$. The same initial conditions are used for all three
communication topologies.

For the star and cyclic networks, all sensing nodes employ the common
Hurwitz matrix
\begin{equation*}
    H_i=-1.5I_2,
    \qquad
    i=1,\dots,4,
\end{equation*}
together with the uniform adaptation gains
\begin{equation*}
    \gamma_{A_i}=15,
    \qquad
    \gamma_{C_i}=15,
    \qquad
    \gamma_{u_i}=30.
\end{equation*}
In the path topology, only the first sensing node has direct access to
the source. Since the source information must propagate successively
through the remaining nodes, progressively faster observer dynamics are
assigned downstream:
\begin{equation*}
    \begin{aligned}
    H_1&=-2.5I_2,
    &
    H_2&=-4I_2,
    \\
    H_3&=-6I_2,
    &
    H_4&=-8I_2.
    \end{aligned}
\end{equation*}
The corresponding adaptation gains are
\begin{equation*}
    \begin{aligned}
    [\gamma_{A_1},\gamma_{A_2},\gamma_{A_3},\gamma_{A_4}]
    &=
    [15,19,23,27],\\
    [\gamma_{C_1},\gamma_{C_2},\gamma_{C_3},\gamma_{C_4}]
    &=
    [15,19,23,27],\\
    [\gamma_{u_1},\gamma_{u_2},\gamma_{u_3},\gamma_{u_4}]
    &=
    [30,39,49,61].
    \end{aligned}
\end{equation*}
The progressively faster downstream dynamics reduce the phase lag
introduced by the sequential propagation of information along the path.

Each sensing node implements \eqref{eq:Problem:sensor} using the
locally aggregated state in \eqref{eq:Problem:z}. In particular, the
nonlinear mapping is evaluated at $z_i$ rather than at the unavailable
source state $x_0$. The estimates $\hat A_i$, $\hat C_i$, and the local
input $u_i$ are updated according to \eqref{eq:AdpId:updates}. Unlike
the discrete-time implementation, the continuous-time adaptive laws do
not require a normalization factor.

The state and adaptive differential equations are integrated over
$t\in[0,100]$ using \texttt{ode45}. The relative and absolute
integration tolerances are set to $10^{-8}$ and $10^{-10}$,
respectively, with a maximum integration step of $0.005$.

\subsection{Nominal Case}

\begin{figure*}[h!]
    \centering
    \subfloat[\label{F3a} Star--$\mathcal{G}_{\mathrm{s}}$]{%
    \includegraphics[width=.315\linewidth]{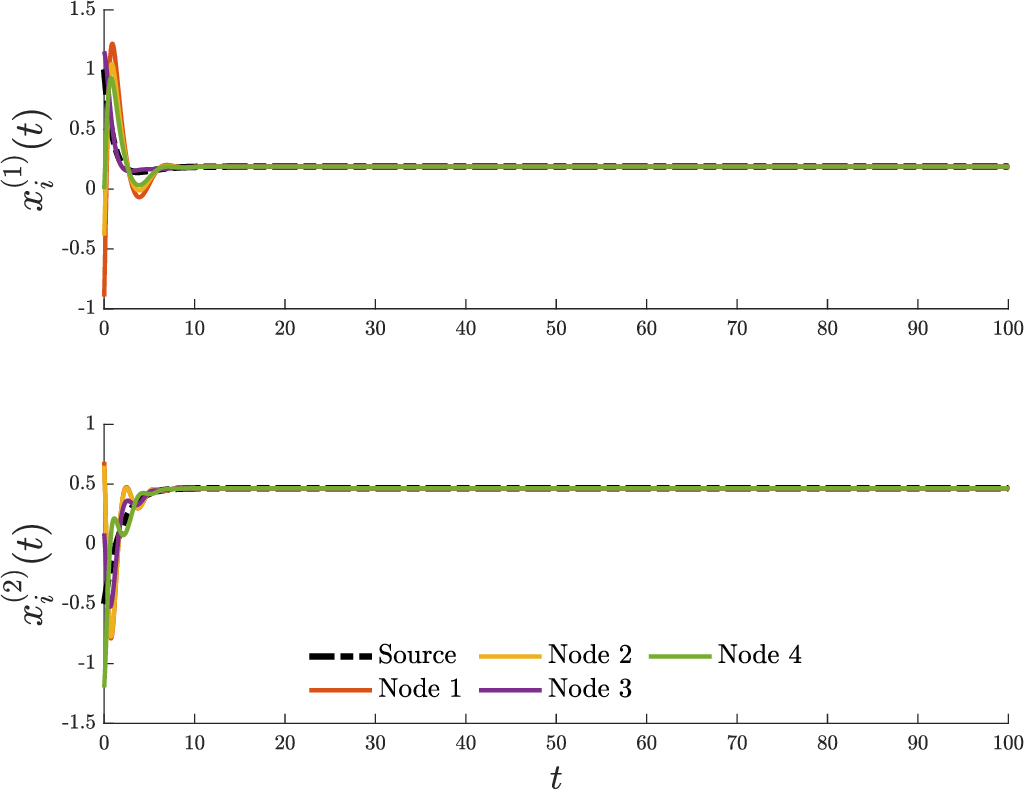}}\quad
    \subfloat[\label{F3b} Cyclic--$\mathcal{G}_{\mathrm{c}}$]{%
    \includegraphics[width=.315\linewidth]{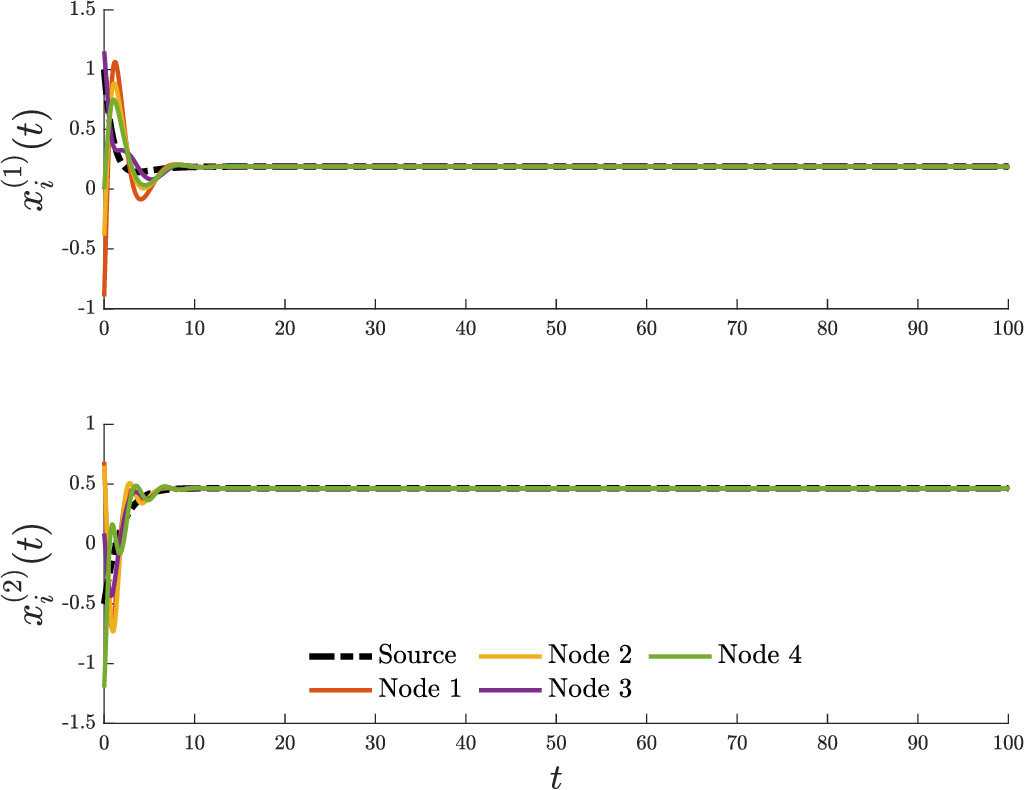}}\quad
    \subfloat[\label{F3c} Path--$\mathcal{G}_{\mathrm{p}}$]{%
    \includegraphics[width=.315\linewidth]{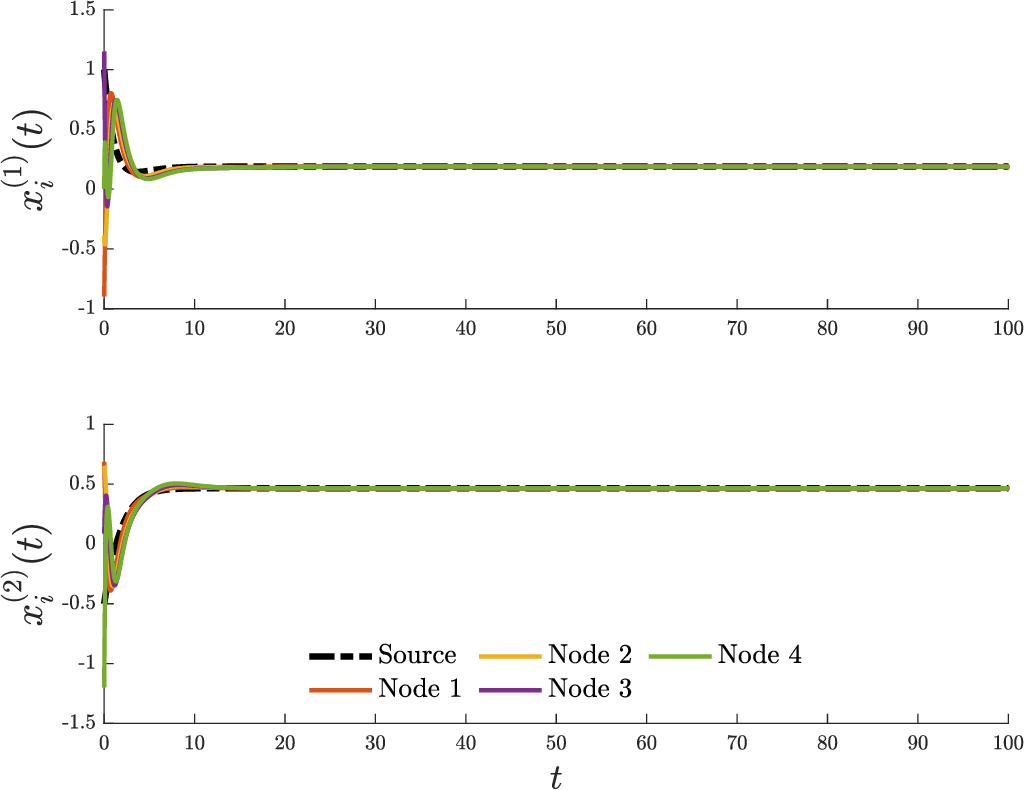}}
    \caption{State trajectories in the nominal case for the three communication topologies: (a) star, (b) cyclic, and (c) path.}
    \label{F3}
\end{figure*}

Figure~\ref{F3} presents the source and sensing-node trajectories in the
nominal case for the star, cyclic, and path topologies. The black
dash-dotted curves represent the source states, while the colored curves
correspond to the four sensing nodes.

For the star topology in Fig.~\ref{F3a}, every sensing node directly
receives the source state. The initial estimation errors are therefore
corrected rapidly, and the four node trajectories approach the source
after a short adaptive transient. The nearly overlapping node
trajectories reflect the identical information pattern and common
observer parameters used by the sensing nodes.

Figure~\ref{F3b} shows the nominal response under the cyclic topology.
Each sensing node combines direct source information with state
estimates received from two neighboring nodes. Although this
interconnection produces a slightly different transient response, all
node estimates approach the source trajectory. 
The scheme achieves reconstruction using both direct and neighboring information.

The path topology is shown in Fig.~\ref{F3c}. Only node $1$ receives the
source state directly, while nodes $2$, $3$, and $4$ obtain source
information successively through the directed chain. This limited
communication structure produces larger initial deviations and a more
visible propagation transient. Nevertheless, the progressively faster
matrices $H_i$ and node-dependent adaptation gains compensate for the
accumulated delay, allowing all sensing nodes to approach the source
trajectory.

\begin{figure*}[h!]
    \centering
    \subfloat[\label{F5a} Adaptive parameter errors]{%
    \includegraphics[width=.315\linewidth]{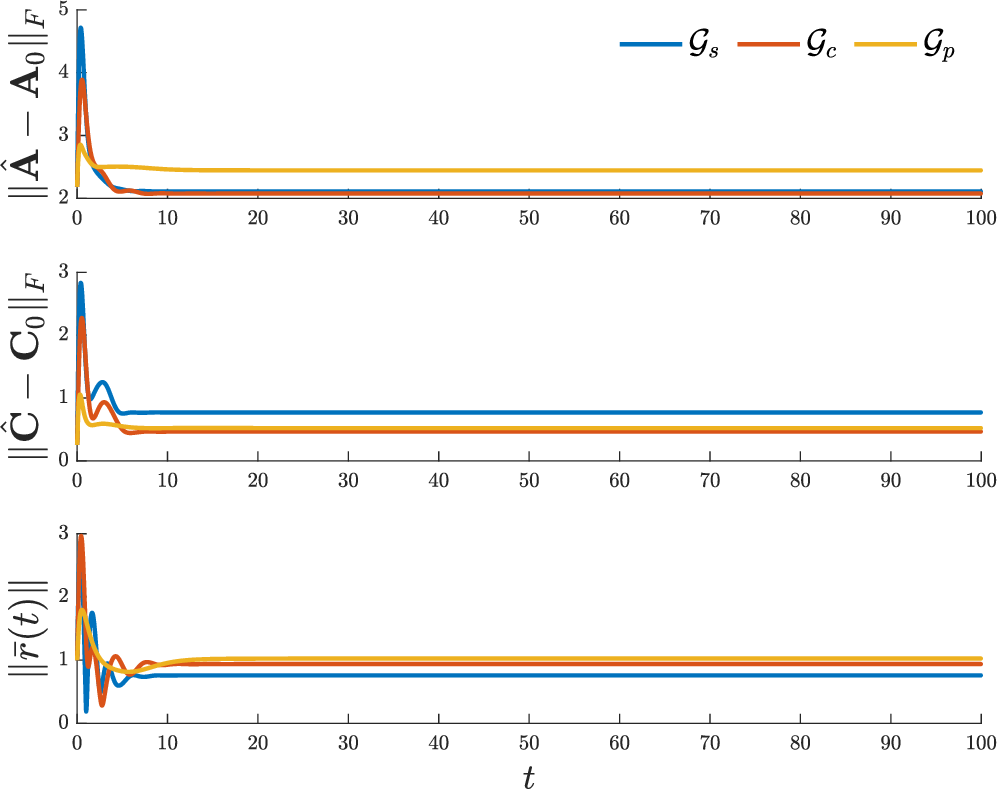}}\quad
    \subfloat[\label{F5b} Network error measures]{%
    \includegraphics[width=.315\linewidth]{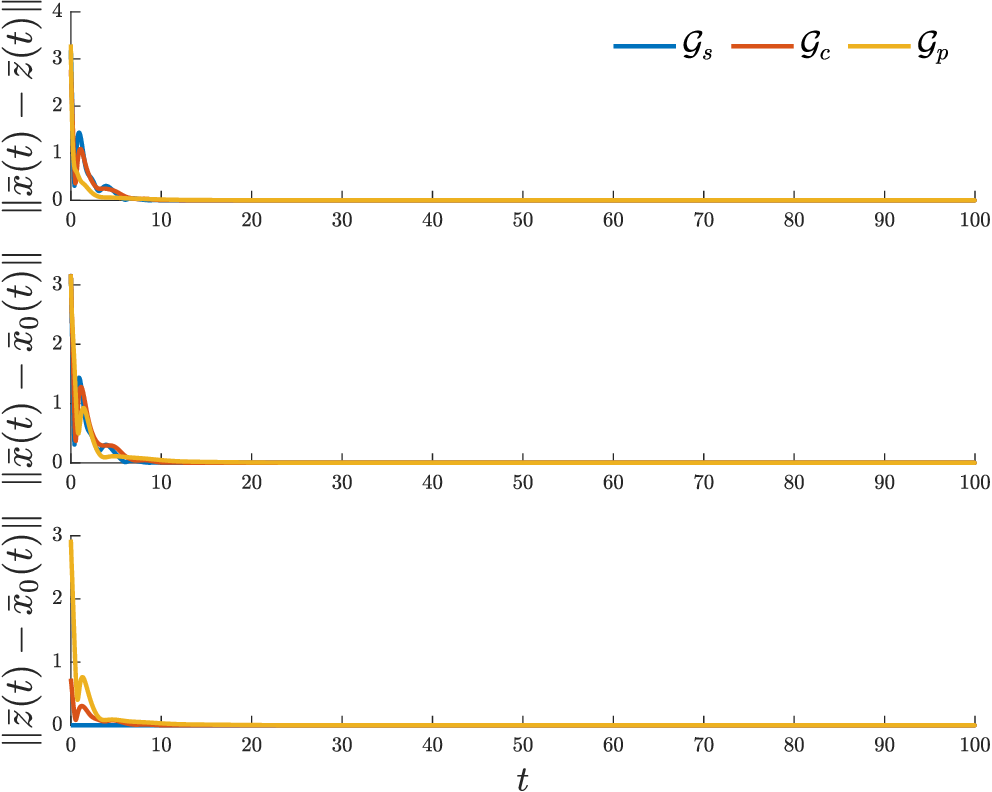}}\quad
    \subfloat[\label{F5c} Nonlinear mapping mismatch]{%
    \includegraphics[width=.315\linewidth]{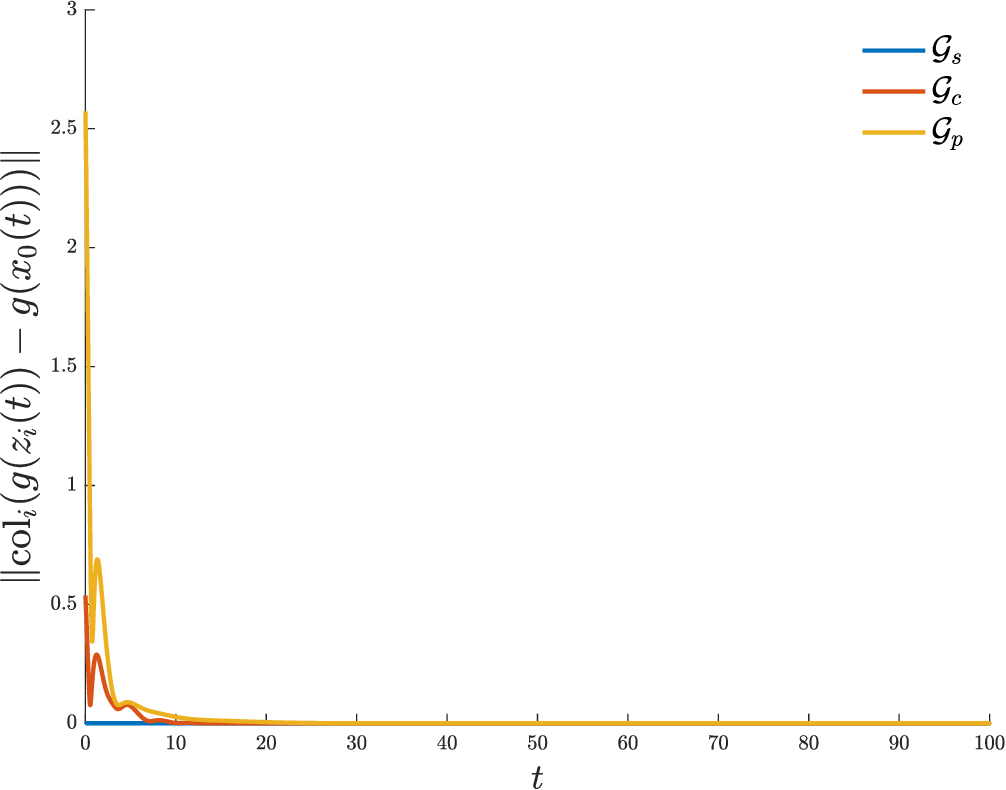}}
    \caption{Error responses in the nominal case: (a) adaptive parameter errors, (b) network estimation and source-tracking errors, and (c) nonlinear mapping mismatch.}
    \label{F5}
\end{figure*}

The nominal error responses are reported in Fig.~\ref{F5}. The adaptive
parameter errors in Fig.~\ref{F5a} remain bounded for all three
topologies. Their largest variations occur during the initial
adaptation period, when the node states differ most significantly from
the source and the network-weighted error directions $\zeta_i$ are
largest. As the state reconstruction errors decrease, the adaptive updates
diminish and the parameter trajectories settle.

The estimates of $A_0$ and $C_0$ remain bounded in the simulations,
in agreement with Theorem~\ref{thm:AdpId:CL}(ii). As discussed in
Remark~\ref{rem:PE}, convergence to the true parameter values generally
requires appropriate excitation and identifiability conditions on the
combined regressors $z_i$ and $g(z_i)$.

Figure~\ref{F5b} shows the network-level error measures
\begin{equation*}
    \|\Bar x-\Bar z\|,
    \qquad
    \|\Bar x-\Bar x_0\|,
    \qquad
    \|\Bar z-\Bar x_0\|.
\end{equation*}
The first quantity is the stacked local estimation error $\|\Bar e\|$,
the second is the source-tracking error $\|\Bar\epsilon\|$, and the
third measures the discrepancy between the aggregated node states and
the replicated source state.

All three error measures decrease toward zero in the nominal case.
This behavior agrees with Theorem~\ref{thm:AdpId:CL}, which establishes
$\Bar e(t)\to0$ and $\Bar\epsilon(t)\to0$ when
$\delta_0(t)\equiv0$ and $\dot u_0(t)\equiv0$. The convergence of
$\Bar z-\Bar x_0$ follows from
\eqref{eq:AdpId:z_tracking_relation}. The differences among the initial
transients reflect the distinct propagation patterns of the three
communication topologies, but do not affect the asymptotic
reconstruction result.

The nonlinear mapping mismatch is displayed in Fig.~\ref{F5c}. This
quantity is defined as
\begin{equation*}
    \left\|
        \operatorname{col}_{i}
        \bigl(g(z_i)-g(x_0)\bigr)
    \right\|.
\end{equation*}
The mismatch decreases toward zero as the aggregated states $z_i$
approach the source state. This observation agrees with the Lipschitz
estimate in \eqref{eq:AdpId:g_mismatch_bound}, which bounds the
nonlinear mismatch directly in terms of $\|\Bar e\|$. Hence, the
locally evaluated nonlinear terms become consistent with the source
nonlinearity as the estimation errors vanish.

Overall, the nominal results illustrate the conclusions of
Theorem~\ref{thm:AdpId:CL}: the estimation and source-tracking errors
approach zero, the adaptive variables remain bounded, and the rates of
the adaptive updates decrease as the closed-loop system approaches its
limiting behavior.

\subsection{ISS Case}

\begin{figure*}[h!]
    \centering
    \subfloat[\label{F4a} Star--$\mathcal{G}_{\mathrm{s}}$]{%
    \includegraphics[width=.315\linewidth]{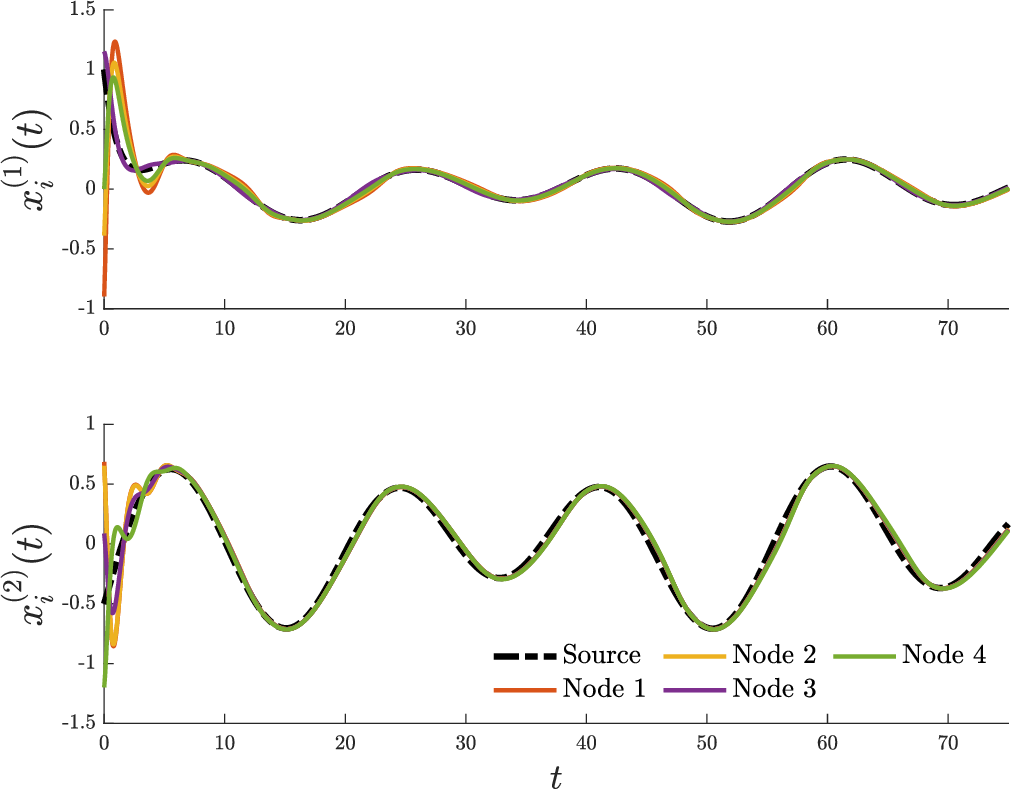}}\quad
    \subfloat[\label{F4b} Cyclic--$\mathcal{G}_{\mathrm{c}}$]{%
    \includegraphics[width=.315\linewidth]{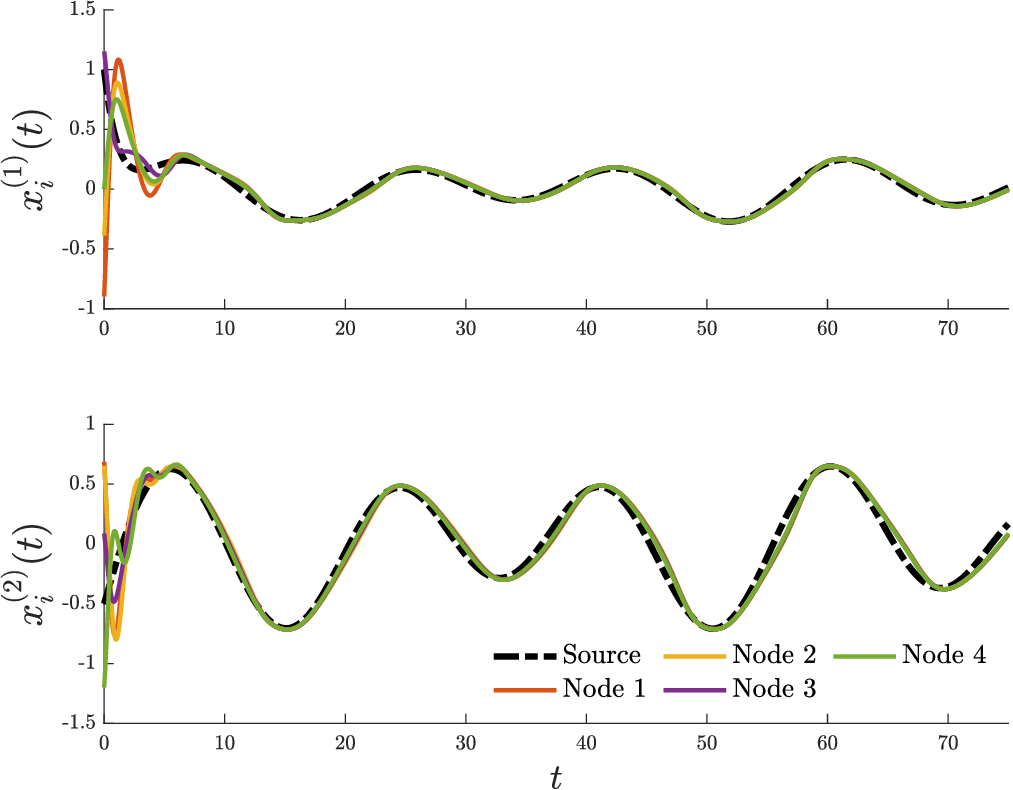}}\quad
    \subfloat[\label{F4c} Path--$\mathcal{G}_{\mathrm{p}}$]{%
    \includegraphics[width=.315\linewidth]{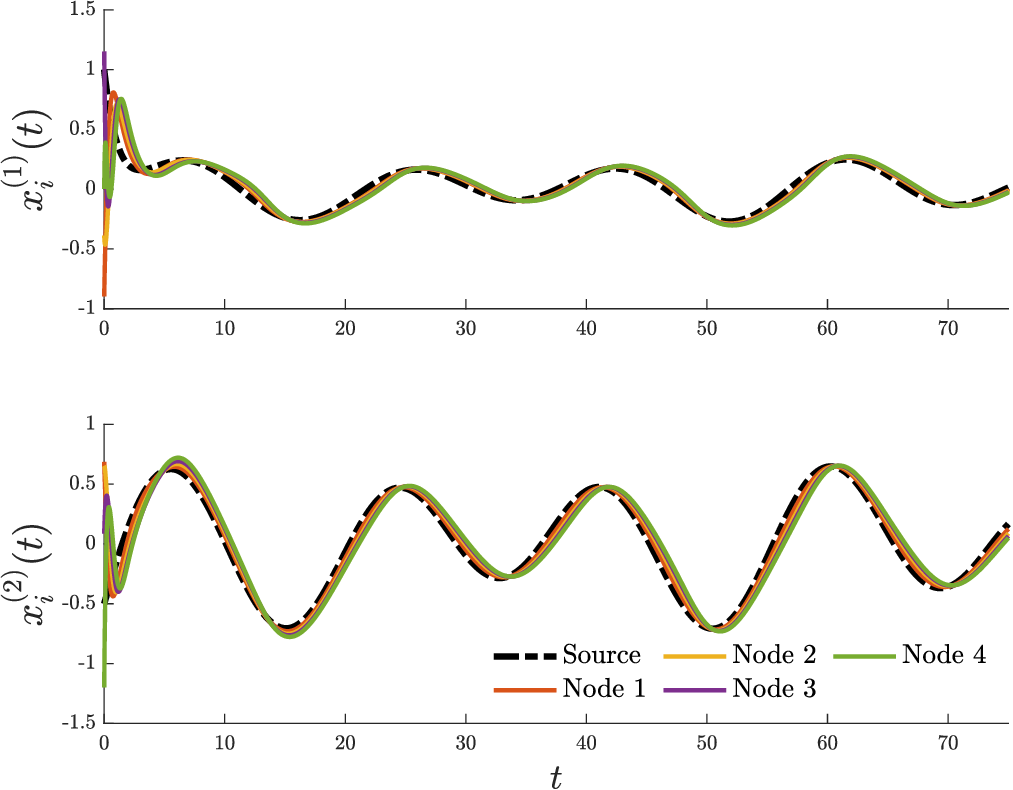}}
    \caption{State trajectories in the ISS case for the three communication topologies: (a) star, (b) cyclic, and (c) path.}
    \label{F4}
\end{figure*}

Figure~\ref{F4} presents the state trajectories in the ISS case, where
the unknown source input varies continuously and the source is affected
by a bounded disturbance.

The star-topology response is shown in Fig.~\ref{F4a}. Because every
sensing node receives direct source information, the four node
trajectories remain close to the source despite the persistent
perturbations. Small residual deviations are visible because the local
inputs continuously adapt to the time-varying source input, but the
tracking errors remain bounded throughout the simulation.

Figure~\ref{F4b} presents the cyclic-topology response. Each sensing
node combines partial source information with neighboring estimates.
The node trajectories continue to follow the source closely and remain
bounded in the presence of the time-varying input and disturbance. The
result indicates that the cyclic interconnection does not produce an
unbounded amplification of the persistent perturbations.

\begin{figure*}[h!]
    \centering
    \subfloat[\label{F6a} Adaptive parameter errors]{%
    \includegraphics[width=.315\linewidth]{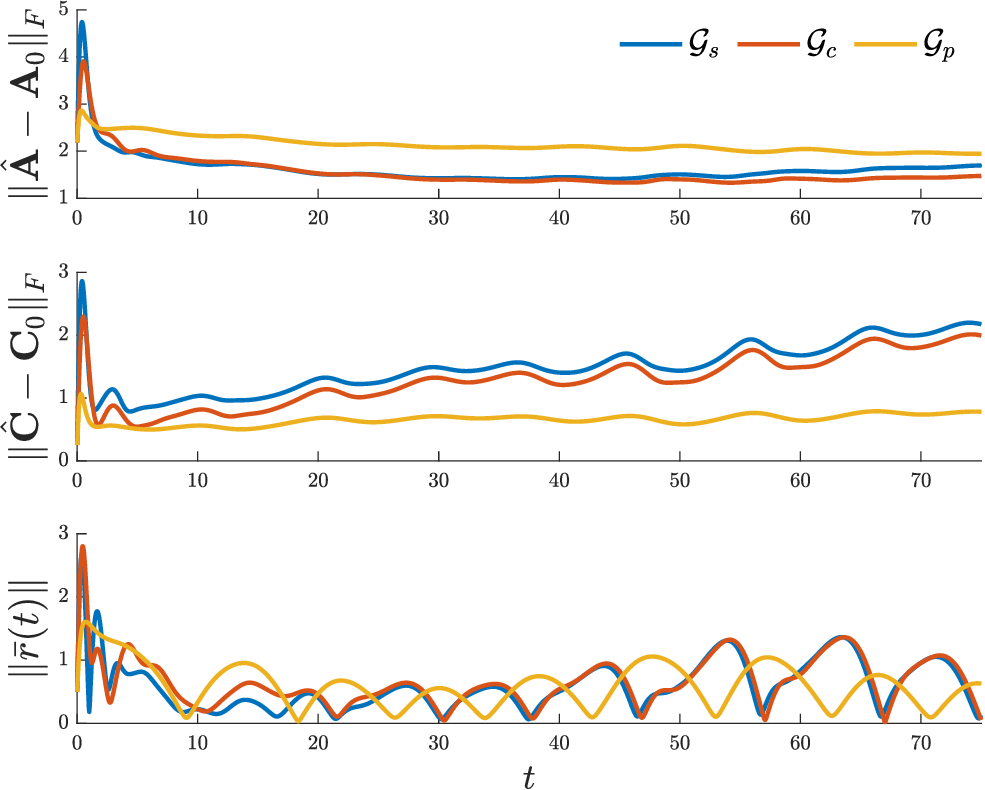}}\quad
    \subfloat[\label{F6b} Network error measures]{%
    \includegraphics[width=.315\linewidth]{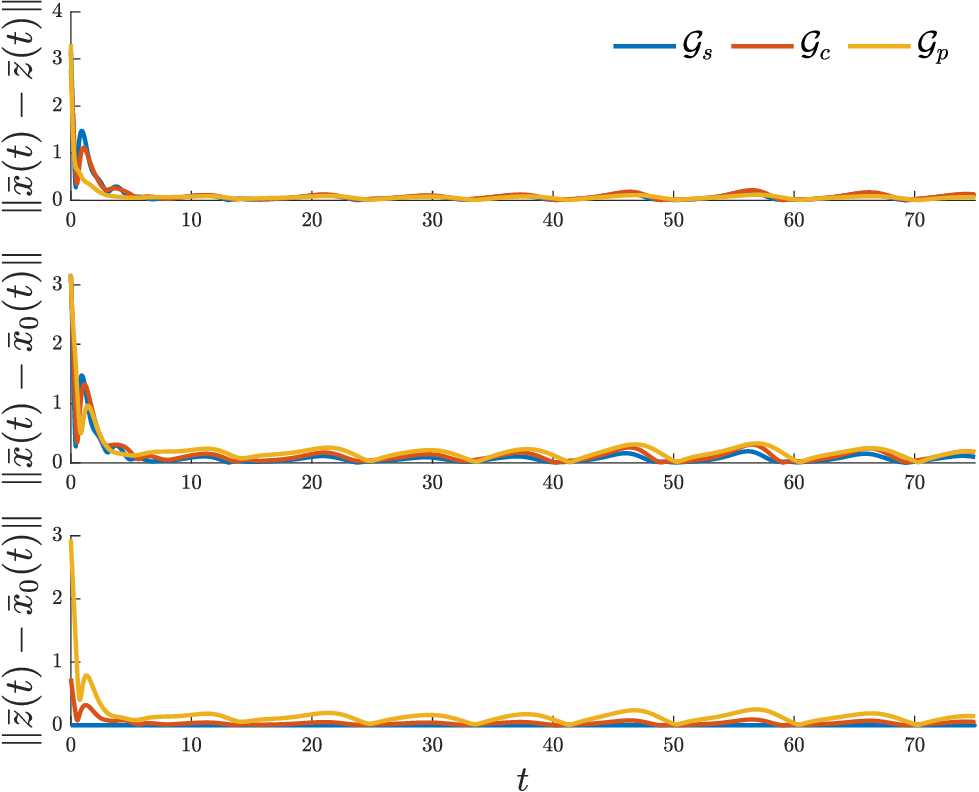}}\quad
    \subfloat[\label{F6c} Nonlinear mapping mismatch]{%
    \includegraphics[width=.315\linewidth]{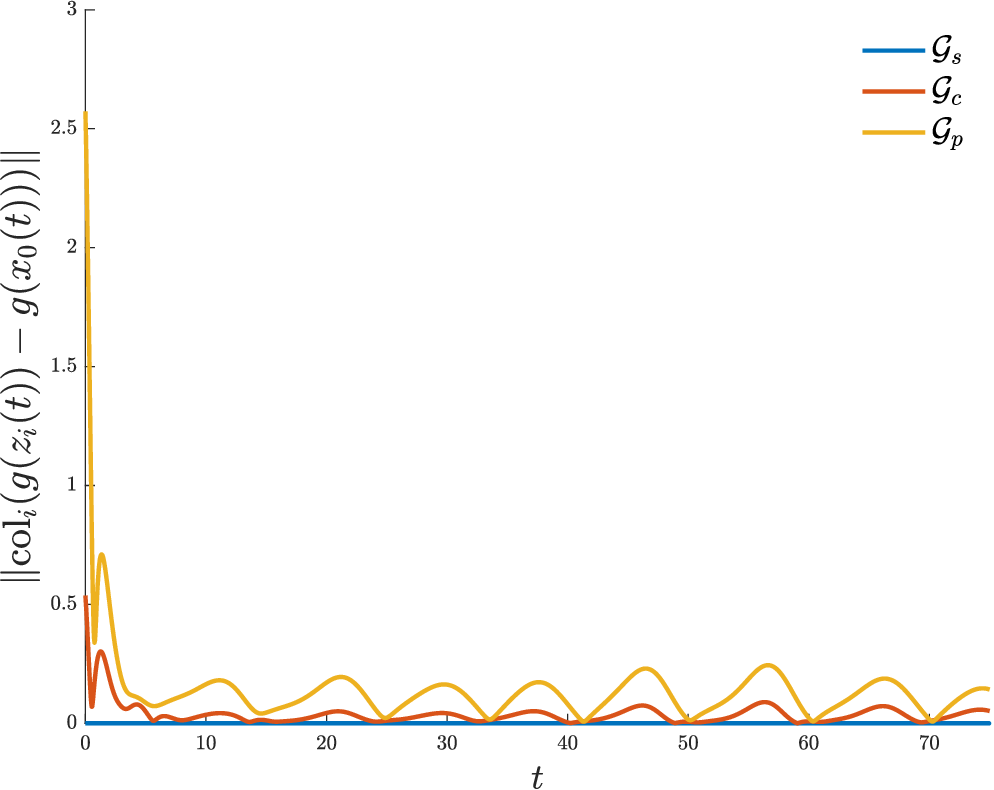}}
    \caption{Error responses in the ISS case: (a) adaptive parameter errors, (b) network estimation and source-tracking errors, and (c) nonlinear mapping mismatch.}
    \label{F6}
\end{figure*}

The path-topology response is shown in Fig.~\ref{F4c}. Since only the
first node receives direct source information, the time-varying source
signal must propagate through several dynamical layers before reaching
the final node. Small phase differences are therefore more visible,
particularly among the downstream nodes. The progressively faster
observer matrices and larger adaptation gains reduce this accumulated
lag, and all node trajectories remain bounded while continuing to
follow the source.

Figure~\ref{F6a} presents the adaptive parameter errors in the ISS
case. The parameter trajectories remain bounded across the entire
simulation, although persistent changes are expected because the
adaptive inputs must continuously respond to $\dot u_0(t)\neq0$. In
particular, the input mismatch
$\Bar r=\Bar u-\Bar u_0$ generally approaches a bounded time-varying
response rather than converging to zero. The bounded parameter
trajectories also ensure that the adaptive predictor mismatch remains
bounded over the simulated interval.

The network estimation and source-tracking errors are shown in
Fig.~\ref{F6b}. In contrast to the nominal case, the errors generally
approach bounded neighborhoods of the origin rather than converging
exactly to zero. The residual oscillations are induced by the
time-varying source input and external disturbance. Their amplitudes
remain uniformly bounded, which is consistent with the ISS robustness
result of Corollary~\ref{cor:AdpId:ISS}. The different transient and
ultimate error levels reflect how the bounded perturbations propagate
through the corresponding communication structures.

Figure~\ref{F6c} shows the nonlinear mapping mismatch in the ISS case.
The mismatch remains bounded and follows the behavior of the
aggregated-state error. This is consistent with
\eqref{eq:AdpId:g_mismatch_bound}, which ensures that the nonlinear
mapping mismatch is controlled by the estimation error. Consequently,
the locally evaluated nonlinear term does not introduce an independent
unbounded perturbation into the error dynamics.

Taken together, the ISS results show that the proposed adaptive
observers maintain bounded state, parameter, and nonlinear mismatch
responses under persistent bounded perturbations. Exact convergence is
not expected in this setting because both $\dot u_0(t)$ and
$\delta_0(t)$ remain nonzero. Instead, the observed behavior agrees
with the ultimate boundedness characterized in
Corollary~\ref{cor:AdpId:ISS}.

\subsection{Comparison of the Coupling-Operator Conditions}

\begin{table*}[t!]
    \centering
    \caption{Numerical comparison of the Hurwitz stability conditions for the coupling operator $\mathcal H$.}
    \label{tab:H_numerical}
    \begin{tabular}{lcccccc}
        \toprule
        Topology &
        $\alpha(\mathcal H)$ &
        Prop.~1 LHS &
        Prop.~1 Margin &
        Lemma~1 Margin &
        Cor.~1 Margin &
        Best $\varepsilon$ \\
        \midrule
        Star
        & $-1.5000$
        & $-1.5000$
        & $1.500$
        & $0.750$
        & $0.750$
        & $3.090\times10^{-7}$ \\
        
        Cyclic
        & $-1.2675$
        & $-0.7929$
        & $0.7929$
        & $0.7236$
        & $0.6903$
        & $1.818\times10^{-1}$ \\
        
        Path
        & $-2.5000$
        & $-0.5223$
        & $0.5223$
        & $1.644$
        & $1.628$
        & $1.279\times10^{-1}$ \\
        \bottomrule
    \end{tabular}
\end{table*}

\begin{table}[t!]
    \centering
    \caption{Comparison of the Hurwitz stability certificates for the coupling operator $\mathcal H$.}
    \label{tab:H_summary}
    \begin{tabular}{lcccc}
        \toprule
        Topology &
        $\alpha(\mathcal H)<0$ &
        Proposition~1 &
        Lemma~1 &
        Corollary~1 \\
        \midrule
        Star
        & $\checkmark$
        & $\checkmark$
        & $\checkmark$
        & $\checkmark$ \\
        
        Cyclic
        & $\checkmark$
        & $\checkmark$
        & $\checkmark$
        & $\checkmark$ \\
        
        Path
        & $\checkmark$
        & $\checkmark$
        & $\checkmark$
        & $\checkmark$ \\
        \bottomrule
    \end{tabular}
\end{table}

We next compare the three Hurwitz stability certificates developed for
the coupling operator $\mathcal H$. The source matrix is decomposed as
$A_0=A_\ast+\Delta A$, where
\begin{equation*}
    A_\ast
    =
    \begin{bmatrix}
        -0.75 & 0.28\\
        -0.18 & -0.62
    \end{bmatrix},
\end{equation*}
and the uncertainty radius is selected as
\begin{equation*}
    a
    =
    \|A_0-A_\ast\|_2.
\end{equation*}
The structured certificates use $P_n=I_2$. The matrix $P_m\succ0$ and,
for the robust condition, the scalar $\varepsilon>0$ are determined
through a multistart numerical search. A positive feasibility margin
indicates that the corresponding sufficient condition is satisfied.

Table~\ref{tab:H_numerical} reports the spectral abscissa
\begin{equation*}
    \alpha(\mathcal H)
    \coloneqq
    \max_{\lambda\in\lambda(\mathcal H)}
    \Re\{\lambda\},
\end{equation*}
together with the numerical results associated with
Proposition~\ref{prop:NL:H_hurwitz},
Lemma~\ref{lem:NL:LMI_H}, and
Corollary~\ref{cor:NL:robust_H}. The spectral abscissa are
$-1.5000$, $-1.2675$, and $-2.5000$ for the star, cyclic, and path
topologies, respectively. Thus, the coupling operator is Hurwitz in all
three cases.

The logarithmic-norm condition in
Proposition~\ref{prop:NL:H_hurwitz} is satisfied for every topology.
Its left-hand side is $-1.5000$ for the star network, $-0.7929$ for the
cyclic network, and $-0.5223$ for the path network. The corresponding
positive margins are $1.500$, $0.7929$, and $0.5223$. The path network
has the smallest proposition margin, reflecting the effect of its
asymmetric and sequential information flow. Nevertheless, the
sufficient condition remains strictly satisfied.

The structured Lyapunov condition of
Lemma~\ref{lem:NL:LMI_H} produces positive feasibility margins of
$0.750$, $0.7236$, and $1.644$ for the star, cyclic, and path
topologies, respectively. These values confirm the existence of
structured quadratic stability certificates for the selected coupling
operators.

The robust condition in Corollary~\ref{cor:NL:robust_H} also gives
positive margins for all three topologies. The corresponding margins
are $0.750$, $0.6903$, and $1.628$, with the selected values of
$\varepsilon$ equal to
$3.090\times10^{-7}$,
$1.818\times10^{-1}$, and
$1.279\times10^{-1}$, respectively.

For the star topology, $\mathbb A_m=0$, and therefore the uncertainty
in $A_0$ does not enter the coupling operator through the
neighbor-coupling term. This explains why the nominal and robust
structured margins coincide and why the numerically selected
$\varepsilon$ is close to zero. For the cyclic and path networks, the
uncertain source dynamics enter through the sensing-to-sensing
coupling, and the robust certificate explicitly accounts for their
effect.

\begin{figure}[t!]
    \centering
    \includegraphics[width=.45\linewidth]{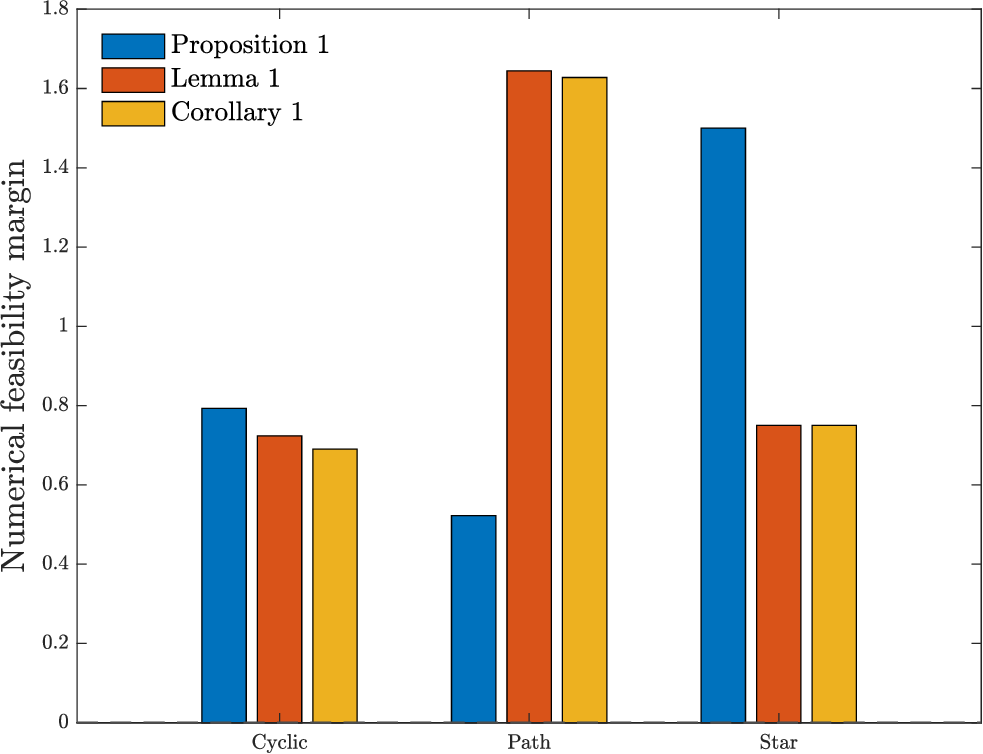}
    \caption{Numerical feasibility margins of Proposition~\ref{prop:NL:H_hurwitz}, Lemma~\ref{lem:NL:LMI_H}, and Corollary~\ref{cor:NL:robust_H} for the continuous-time coupling operator $\mathcal H$.}
    \label{F9}
\end{figure}

Table~\ref{tab:H_summary} summarizes the pass/fail results. The direct
Hurwitz test, Proposition~\ref{prop:NL:H_hurwitz},
Lemma~\ref{lem:NL:LMI_H}, and
Corollary~\ref{cor:NL:robust_H} certify all three communication
topologies. Figure~\ref{F9} provides a visual comparison of the
feasibility margins. All reported margins are positive, confirming
that the selected continuous-time observer and network parameters
satisfy the proposed stability conditions.

The comparison highlights the complementary roles of the three
conditions. Proposition~\ref{prop:NL:H_hurwitz} provides a direct
analytical test based on the logarithmic norm and induced matrix norm.
Lemma~\ref{lem:NL:LMI_H} retains the full matrix structure of the
coupling operator through a structured Lyapunov function.
Corollary~\ref{cor:NL:robust_H} extends the structured certificate to
all source matrices contained in the prescribed norm-bounded
uncertainty set.

\subsection{Computational Scalability}

Finally, we evaluate the computational scalability of the proposed
continuous-time adaptive identification method for
$m\in\{50,100,250,500\}$ sensing nodes. A direct comparison based on an
adaptive-step ODE solver could be misleading because the number of
internal integration steps depends on the error tolerances, stiffness,
topology, and simulated trajectory. To isolate the online cost of
evaluating the distributed dynamics, the complexity benchmark therefore
uses a fixed-step fourth-order Runge--Kutta method.

The integration step is set to $0.02$, and every experiment is
performed for $K=300$ RK4 steps, corresponding to the horizon $T=6$.
Since each RK4 step requires four evaluations of the closed-loop vector
field, every experiment uses exactly $1200$ vector-field evaluations.
The graph matrices are stored in sparse form, and the same state
dimension, nonlinear mapping, and adaptive update structure are used
for all network sizes. Each experiment is repeated three times, and
the median runtime is reported to reduce timing variability.

Table~\ref{tab:timing} reports the total runtime for the three
communication topologies. For the star topology, the runtime increases
from $0.2546$ seconds at $m=50$ to $1.7389$ seconds at $m=500$. For the
cyclic topology, it increases from $0.2100$ to $1.7271$ seconds, while
the path-topology runtime increases from $0.2016$ to $1.7088$ seconds.
The runtimes remain similar across the three sparse communication
structures for each fixed network size.

To quantify the observed scaling, the measured runtimes are fitted to
the model
\begin{equation*}
    t_{\mathrm{tot}}
    \approx
    c\,m^q.
\end{equation*}
The resulting empirical exponents are
$q=0.848$ for the star topology,
$q=0.923$ for the cyclic topology, and
$q=0.932$ for the path topology. These values indicate approximately
linear growth with respect to the number of sensing nodes. The slightly
sublinear exponents are attributable to fixed implementation overhead,
memory allocation, vectorization, and cache effects, which have a
larger relative influence for smaller networks.

The average computational cost per node and vector-field evaluation
also remains nearly constant as the network size increases. At
$m=500$, the measured costs are approximately $2.90$, $2.88$, and
$2.85$ microseconds per node and vector-field evaluation for the star,
cyclic, and path topologies, respectively. This behavior is consistent
with the sparse graph representations and local adaptive updates, whose
online costs grow proportionally with the number of sensing nodes for
bounded-degree networks.

Although continuous-time integration requires multiple evaluations of
the vector field during each integration step, the cost of each
evaluation scales approximately linearly with the network size. The
results therefore demonstrate that the proposed distributed adaptive
identification method remains computationally practical for networks
containing several hundred sensing nodes.

\begin{table*}[t!]
    \centering
    \caption{Runtime $t_{\mathrm{tot}}$ (s) and empirical scaling exponent for the fixed-step RK4 complexity benchmark. Each experiment uses $1200$ vector-field evaluations.}
    \label{tab:timing}
    \begin{tabular}{lccccc}
        \toprule
        Topology &
        $m=50$ &
        $m=100$ &
        $m=250$ &
        $m=500$ &
        Empirical order \\
        \midrule
        Star
        & $0.2546$
        & $0.3741$
        & $0.8819$
        & $1.7389$
        & $\mathcal O(m^{0.848})$ \\
        
        Cyclic
        & $0.2100$
        & $0.3666$
        & $0.8958$
        & $1.7271$
        & $\mathcal O(m^{0.923})$ \\
        
        Path
        & $0.2016$
        & $0.3647$
        & $0.8757$
        & $1.7088$
        & $\mathcal O(m^{0.932})$ \\
        \bottomrule
    \end{tabular}
\end{table*}

\section{Conclusion and Future Work}
\label{sec:Conclusion}

This paper developed a continuous-time distributed adaptive
identification framework for unknown nonlinear systems over directed
communication networks. Each sensing node reconstructs the source state
and adapts local estimates of its unknown dynamics using only locally
available information and neighboring state estimates, without sharing
the unknown source input or the local inputs of other nodes.

Sufficient conditions were established for the Hurwitz stability of the
network coupling operator, including a direct norm-based condition, a
structured Lyapunov certificate, and a robust LMI accounting for
norm-bounded model uncertainty. Unlike the discrete-time formulation,
the continuous-time adaptive laws do not require normalization because
the Lyapunov derivative does not generate quadratic adaptive-update
terms. The stability analysis established boundedness of the adaptive
parameters and asymptotic convergence of the estimation and
source-tracking errors in the unperturbed case. An ISS result was also
provided under bounded predictor mismatch and source disturbances.

Numerical studies over star, cyclic, and path topologies supported the
theoretical results. The nominal simulations demonstrated asymptotic
source reconstruction, while the perturbed simulations showed bounded
tracking errors under a time-varying unknown input and external
disturbance. The proposed stability certificates were verified
numerically for all three topologies. Moreover, the complexity
benchmark demonstrated approximately linear growth in online
computational cost with the number of sensing nodes.

Future work will investigate verifiable cooperative excitation
conditions for exact parameter identification and robust adaptive
mechanisms for persistent disturbances and time-varying inputs.
Extensions to switching topologies, communication delays, output-only
measurements, and experimental networked systems will also be
considered.

\bibliographystyle{IEEEtran}    
\bibliography{EN-Bib.bib}

\newpage

\appendix

\section{Proofs}
\paragraph*{Proof of Proposition~\ref{prop:NL:H_hurwitz}:}
Write the coupling operator as
\begin{equation*}
    \mathcal H
    =
    \Omega\mathbf H
    +
    (\mathbb A_m\otimes I_n)\mathbf A_0.
\end{equation*}
By the subadditivity of the Euclidean logarithmic norm and the bound
$\mu_2(M)\le\|M\|_2$, we have
\begin{align}
    \mu_2(\mathcal H)
    &\le
    \mu_2(\Omega\mathbf H)
    +
    \mu_2\bigl((\mathbb A_m\otimes I_n)\mathbf A_0\bigr)
    \nonumber\\
    &\le
    \mu_2(\Omega\mathbf H)
    +
    \|(\mathbb A_m\otimes I_n)\mathbf A_0\|_2.
    \label{eq:pf:H_log_norm}
\end{align}
Since $\mathbf A_0=I_m\otimes A_0$, the mixed-product property of the
Kronecker product gives
\begin{equation*}
    (\mathbb A_m\otimes I_n)\mathbf A_0
    =
    \mathbb A_m\otimes A_0.
\end{equation*}
Consequently, using $\|A\otimes B\|_2=\|A\|_2\|B\|_2$, we obtain
\begin{equation}\label{eq:pf:H_source_bound}
    \|(\mathbb A_m\otimes I_n)\mathbf A_0\|_2
    =
    \|\mathbb A_m\otimes A_0\|_2
    =
    \|\mathbb A_m\|_2\|A_0\|_2.
\end{equation}
Combining \eqref{eq:pf:H_log_norm} and
\eqref{eq:pf:H_source_bound} yields
\begin{equation*}
    \mu_2(\mathcal H)
    \le
    \mu_2(\Omega\mathbf H)
    +
    \|\mathbb A_m\|_2\|A_0\|_2.
\end{equation*}
Therefore, condition~\eqref{eq:NL:H_hurwitz:cond} implies
$\mu_2(\mathcal H)<0$. Since
\begin{equation*}
    \max_{\lambda\in\lambda(\mathcal H)}
    \Re\{\lambda\}
    \le
    \mu_2(\mathcal H),
\end{equation*}
every eigenvalue of $\mathcal H$ has strictly negative real part.
Hence, $\mathcal H$ is Hurwitz.
\hfill$\square$

\paragraph*{Proof of Lemma~\ref{lem:NL:LMI_H}:}
Since $P_m\succ0$ and $P_n\succ0$, their Kronecker product satisfies
\begin{equation*}
    P_{mn}
    =
    P_m\otimes P_n
    \succ0.
\end{equation*}
Consider the quadratic function
\begin{equation*}
    V(\Bar e)
    =
    \Bar e^\top P_{mn}\Bar e.
\end{equation*}
Along the nominal error dynamics
$\dot{\Bar e}=\mathcal H\Bar e$, its derivative is
\begin{align}
    \dot V(\Bar e)
    &=
    \dot{\Bar e}^{\top}P_{mn}\Bar e
    +
    \Bar e^\top P_{mn}\dot{\Bar e}
    \nonumber\\
    &=
    \Bar e^\top
    \left(
        \mathcal H^\top P_{mn}
        +
        P_{mn}\mathcal H
    \right)
    \Bar e.
\end{align}
Condition~\eqref{eq:NL:structured_LMI} implies
$\dot V(\Bar e)<0$ for every $\Bar e\neq0$. Thus, the origin of
$\dot{\Bar e}=\mathcal H\Bar e$ is exponentially stable. By the
continuous-time Lyapunov stability theorem, this is equivalent to
$\mathcal H$ being Hurwitz.
\hfill$\square$

\paragraph*{Proof of Corollary~\ref{cor:NL:robust_H}:}
Let
\begin{equation*}
    d
    \coloneqq
    a\|\mathbb A_m\|_2.
\end{equation*}
Because
$\Delta\mathcal H
=(\mathbb A_m\otimes I_n)(I_m\otimes\Delta A)$, the mixed-product
property and the Kronecker-product norm identity give
\begin{align}
    \|\Delta\mathcal H\|_2
    =
    \|\mathbb A_m\otimes\Delta A\|_2
    =
    \|\mathbb A_m\|_2\|\Delta A\|_2
    \le
    a\|\mathbb A_m\|_2
    =
    d.
    \label{eq:NL:robust_deltaH}
\end{align}
Consequently,
\begin{equation}\label{eq:NL:robust_delta_bound}
    \Delta\mathcal H^\top\Delta\mathcal H
    \preceq
    d^2I_{mn}.
\end{equation}

Applying the Schur complement to
\eqref{eq:NL:robust_LMI}, define
\begin{equation*}
    X
    \coloneqq
    \mathcal H_\ast^\top P_{mn}
    +
    P_{mn}\mathcal H_\ast
    +
    \frac{d^2}{\varepsilon}I_{mn}.
\end{equation*}
Since $\varepsilon>0$, the lower-right block satisfies
\begin{equation*}
    -\frac{1}{\varepsilon}I_{mn}\prec0,
    \qquad
    \left(
        -\frac{1}{\varepsilon}I_{mn}
    \right)^{-1}
    =
    -\varepsilon I_{mn}.
\end{equation*}
The Schur-complement theorem therefore gives
\begin{align*}
    \begin{bmatrix}
        X & P_{mn}\\
        P_{mn} & -\dfrac{1}{\varepsilon}I_{mn}
    \end{bmatrix}
    \prec0
    \quad\Longleftrightarrow\quad
    X
    -
    P_{mn}
    \left(
        -\frac{1}{\varepsilon}I_{mn}
    \right)^{-1}
    P_{mn}
    \prec0.
\end{align*}
Since $P_{mn}=P_{mn}^\top$, the latter inequality becomes
\begin{equation}\label{eq:NL:robust_schur}
    \mathcal H_\ast^\top P_{mn}
    +
    P_{mn}\mathcal H_\ast
    +
    \varepsilon P_{mn}^2
    +
    \frac{d^2}{\varepsilon}I_{mn}
    \prec0.
\end{equation}
Thus, the robust LMI requires the nominal Lyapunov matrix to remain
negative definite after including the two terms that will be used to
dominate the uncertain cross term.

Indeed, for any admissible $\Delta\mathcal H$, the matrix
\begin{equation*}
    \left(
        \sqrt{\varepsilon}P_{mn}
        -
        \frac{1}{\sqrt{\varepsilon}}
        \Delta\mathcal H^\top
    \right)
    \left(
        \sqrt{\varepsilon}P_{mn}
        -
        \frac{1}{\sqrt{\varepsilon}}
        \Delta\mathcal H^\top
    \right)^\top
\end{equation*}
is positive semidefinite. Expanding this expression and using
$P_{mn}=P_{mn}^\top$ gives
\begin{equation*}
    \varepsilon P_{mn}^2
    -
    P_{mn}\Delta\mathcal H
    -
    \Delta\mathcal H^\top P_{mn}
    +
    \frac{1}{\varepsilon}
    \Delta\mathcal H^\top\Delta\mathcal H
    \succeq0.
\end{equation*}
Rearranging yields the weighted Young inequality
\begin{equation}\label{eq:NL:robust_young}
    P_{mn}\Delta\mathcal H
    +
    \Delta\mathcal H^\top P_{mn}
    \preceq
    \varepsilon P_{mn}^2
    +
    \frac{1}{\varepsilon}
    \Delta\mathcal H^\top\Delta\mathcal H.
\end{equation}
Using the uncertainty bound
$\Delta\mathcal H^\top\Delta\mathcal H\preceq d^2I_{mn}$
in \eqref{eq:NL:robust_young}, we obtain
\begin{equation}\label{eq:NL:robust_cross_bound}
    P_{mn}\Delta\mathcal H
    +
    \Delta\mathcal H^\top P_{mn}
    \preceq
    \varepsilon P_{mn}^2
    +
    \frac{d^2}{\varepsilon}I_{mn}.
\end{equation}

Since
$\mathcal H=\mathcal H_\ast+\Delta\mathcal H$, its Lyapunov matrix
satisfies
\begin{align}
    \mathcal H^\top P_{mn}
    +
    P_{mn}\mathcal H
    &=
    \mathcal H_\ast^\top P_{mn}
    +
    P_{mn}\mathcal H_\ast
    +
    \Delta\mathcal H^\top P_{mn}
    +
    P_{mn}\Delta\mathcal H
    \nonumber\\
    &\preceq
    \mathcal H_\ast^\top P_{mn}
    +
    P_{mn}\mathcal H_\ast
    +
    \varepsilon P_{mn}^2
    +
    \frac{d^2}{\varepsilon}I_{mn}.
\end{align}
The right-hand side is negative definite by
\eqref{eq:NL:robust_schur}. Hence,
\begin{equation*}
    \mathcal H^\top P_{mn}
    +
    P_{mn}\mathcal H
    \prec0
\end{equation*}
for every admissible $\Delta A$.

\paragraph*{Proof of Theorem~\ref{thm:AdpId:CL}:}
By Lemma~\ref{lem:NL:LMI_H}, the structured matrix
\begin{equation*}
    P_{mn}
    =
    P_m\otimes P_n
    =
    \diag\{P_1,\dots,P_m\}
    \succ0
\end{equation*}
satisfies
\begin{equation}\label{eq:pf:Lyap}
    \mathcal H^\top P_{mn}
    +
    P_{mn}\mathcal H
    =
    -Q_{mn},
    \qquad
    Q_{mn}\succ0.
\end{equation}
Recall the estimation-error dynamics
\begin{equation}\label{eq:pf:error_dynamics}
    \dot{\Bar e}
    =
    \mathcal H\Bar e
    +
    \Omega\Bar\eta
    +
    \Omega\mathbf C_0\Delta\mathbf g
    -
    \Omega\Bar\delta_0,
\end{equation}
where $\Bar\eta=\Phi\Bar z+\Psi\mathbf g(\Bar z)+\mathbf B_0\Bar r.$
Recall $\zeta=\Omega^\top P_{mn}\Bar e$ in \eqref{eq:AdpId:zeta}.
Since $P_{mn}=\diag\{P_1,\dots,P_m\}$, the $i$th block of
$\zeta$ is precisely the locally implementable direction
$\zeta_i$ employed in the adaptive laws.
Since
$r_i=u_i-u_0$, its dynamics satisfy
\begin{equation}\label{eq:pf:r_dynamics}
    \dot r_i
    =
    -\gamma_{u_i}B_0^\top\zeta_i
    -
    \dot u_0.
\end{equation}
Equivalently,
\begin{equation}\label{eq:pf:r_dynamics_global}
    \dot{\Bar r}
    =
    -\Gamma_u\mathbf B_0^\top\zeta
    -
    \dot{\Bar u}_0,
\end{equation}
where $\dot{\Bar u}_0\coloneqq\mathbf1_m\otimes\dot u_0.$

Consider the composite Lyapunov function
\begin{equation}\label{eq:pf:V}
    V
    =
    \Bar e^\top P_{mn}\Bar e
    +
    \sum_{i=1}^m
    \frac{1}{\gamma_{A_i}}
    \|\Phi_i\|_F^2
    +
    \sum_{i=1}^m
    \frac{1}{\gamma_{C_i}}
    \|\Psi_i\|_F^2
    +
    \sum_{i=1}^m
    \frac{1}{\gamma_{u_i}}
    \|r_i\|^2.
\end{equation}
Since $P_{mn}\succ0$ and all adaptation gains are positive, $V$ is
positive definite and radially unbounded in
$(\Bar e,\Phi,\Psi,\Bar r)$.

Differentiating \eqref{eq:pf:V} along
\eqref{eq:pf:error_dynamics} gives
\begin{align}
    \dot V
    &=
    \Bar e^\top
    \left(
        \mathcal H^\top P_{mn}
        +
        P_{mn}\mathcal H
    \right)
    \Bar e
    +
    2\Bar e^\top P_{mn}\Omega\Bar\eta
    +
    2\Bar e^\top
    P_{mn}\Omega\mathbf C_0\Delta\mathbf g
    -
    2\Bar e^\top P_{mn}\Omega\Bar\delta_0
    \nonumber\\
    &\quad
    +
    2\sum_{i=1}^m
    \frac{1}{\gamma_{A_i}}
    \tr\left(\Phi_i^\top\dot\Phi_i\right)
    +
    2\sum_{i=1}^m
    \frac{1}{\gamma_{C_i}}
    \tr\left(\Psi_i^\top\dot\Psi_i\right)
    +
    2\sum_{i=1}^m
    \frac{1}{\gamma_{u_i}}
    r_i^\top\dot r_i.
    \label{eq:pf:Vdot_raw}
\end{align}
Using $\zeta=\Omega^\top P_{mn}\Bar e$ and the block-diagonal structure of
$\Bar\eta$, the predictor-mismatch term becomes
\begin{equation}\label{eq:pf:eta_cross}
    2\Bar e^\top P_{mn}\Omega\Bar\eta
    =
    2\sum_{i=1}^m
    \zeta_i^\top
    \left[
        \Phi_i z_i
        +
        \Psi_i g(z_i)
        +
        B_0r_i
    \right].
\end{equation}
The adaptive laws yield
\begin{subequations}
\begin{align}
    \frac{2}{\gamma_{A_i}}
    \tr\left(\Phi_i^\top\dot\Phi_i\right)
    &=
    -2\zeta_i^\top\Phi_i z_i,
    \label{eq:pf:Phi_cancel}\\
    \frac{2}{\gamma_{C_i}}
    \tr\left(\Psi_i^\top\dot\Psi_i\right)
    &=
    -2\zeta_i^\top\Psi_i g(z_i),
    \label{eq:pf:Psi_cancel}\\
    \frac{2}{\gamma_{u_i}}
    r_i^\top\dot r_i
    &=
    -2\zeta_i^\top B_0r_i
    -
    \frac{2}{\gamma_{u_i}}r_i^\top\dot u_0.
    \label{eq:pf:r_cancel}
\end{align}
\end{subequations}
Consequently, all terms involving the adaptive predictor mismatch
$\Bar\eta$ cancel exactly. Unlike the discrete-time case, differentiation
of the parameter-error terms generates no quadratic adaptive-update
terms, and hence no normalization factor is required. Using
\eqref{eq:pf:Lyap}, the exact Lyapunov derivative is
\begin{align}
    \dot V
    &=
    -\Bar e^\top Q_{mn}\Bar e
    +
    2\zeta^\top\mathbf C_0\Delta\mathbf g
    -
    2\zeta^\top\Bar\delta_0
    \nonumber\\
    &\quad
    -
    2\Bar r^\top
    \Gamma_u^{-1}\dot{\Bar u}_0.
    \label{eq:pf:Vdot_exact}
\end{align}

We first establish the general Lyapunov bound in item~(i). From
\eqref{eq:AdpId:g_mismatch_bound},
\begin{equation}\label{eq:pf:g_bound}
    \|\Delta\mathbf g\|
    \le
    \kappa_g\|\Bar e\|,
    \qquad
    \kappa_g
    \coloneqq
    L_g
    \|\mathbb A_m\mathbb L^{-1}\otimes I_n\|_2.
\end{equation}
Moreover, $\zeta=\Omega^\top P_{mn}\Bar e$ gives
\begin{equation}\label{eq:pf:zeta_bound}
    \|\zeta\|
    \le
    \|\Omega^\top P_{mn}\|_2\|\Bar e\|.
\end{equation}
Because $\mathbf C_0=I_m\otimes C_0$, one has
$\|\mathbf C_0\|_2=\|C_0\|_2$. Hence,
\begin{align}
    2\zeta^\top\mathbf C_0\Delta\mathbf g
    &\le
    2\|\zeta\|
    \|\mathbf C_0\|_2
    \|\Delta\mathbf g\|
    \nonumber\\
    &\le
    2\|\Omega^\top P_{mn}\|_2
    \|C_0\|_2
    \kappa_g
    \|\Bar e\|^2.
    \label{eq:pf:nonlinear_bound}
\end{align}
Define
\begin{equation}\label{eq:pf:alpha_0}
    \alpha_0
    \coloneqq
    \lambda_{\min}(Q_{mn})
    -
    2\|\Omega^\top P_{mn}\|_2
    \|C_0\|_2
    L_g
    \|\mathbb A_m\mathbb L^{-1}\otimes I_n\|_2.
\end{equation}
The nonlinear small-gain condition guarantees $\alpha_0>0$.

Next, let
\begin{equation*}
    c_\zeta
    \coloneqq
    \|\Omega^\top P_{mn}\|_2.
\end{equation*}
For any $\varepsilon_\delta>0$, Young's inequality gives
\begin{align}
    -2\zeta^\top\Bar\delta_0
    \le
    2\|\zeta\|\|\Bar\delta_0\|
    &\le
    \varepsilon_\delta\|\zeta\|^2
    +
    \frac{1}{\varepsilon_\delta}
    \|\Bar\delta_0\|^2
    \nonumber\\
    &\le
    \varepsilon_\delta c_\zeta^2\|\Bar e\|^2
    +
    \frac{1}{\varepsilon_\delta}
    \|\Bar\delta_0\|^2.
    \label{eq:pf:delta_bound}
\end{align}
Let
$\underline\gamma_u\coloneqq\min_i\gamma_{u_i}>0$.
Since
$\|\Gamma_u^{-1}\|_2=\underline\gamma_u^{-1}$, for any
$\varepsilon_r>0$,
\begin{align}
    -2\Bar r^\top
    \Gamma_u^{-1}\dot{\Bar u}_0
    &\le
    2\|\Gamma_u^{-1}\|_2
    \|\Bar r\|\|\dot{\Bar u}_0\|
    \nonumber\\
    &\le
    \frac{\varepsilon_r}{\underline\gamma_u}
    \|\Bar r\|^2
    +
    \frac{1}{\varepsilon_r\underline\gamma_u}
    \|\dot{\Bar u}_0\|^2.
    \label{eq:pf:u0_bound}
\end{align}
Choose $\varepsilon_\delta>0$ sufficiently small that
\begin{equation}\label{eq:pf:alpha}
    \alpha
    \coloneqq
    \alpha_0-\varepsilon_\delta c_\zeta^2
    >
    0.
\end{equation}
Substituting
\eqref{eq:pf:nonlinear_bound}--\eqref{eq:pf:u0_bound} into
\eqref{eq:pf:Vdot_exact} yields
\begin{align}
    \dot V
    \le{}&
    -\alpha\|\Bar e\|^2
    +
    c_\delta\|\Bar\delta_0\|^2
    +
    \frac{\varepsilon_r}{\underline\gamma_u}
    \|\Bar r\|^2
    +
    c_u\|\dot{\Bar u}_0\|^2,
    \label{eq:pf:Vdot_general}
\end{align}
where
\begin{equation*}
    c_\delta
    \coloneqq
    \frac{1}{\varepsilon_\delta},
    \qquad
    c_u
    \coloneqq
    \frac{1}{\varepsilon_r\underline\gamma_u}.
\end{equation*}
This proves item~(i).

We next prove item~(ii). Under the nominal conditions
$\Bar\delta_0(t)\equiv\mathbf0_{mn}$ and
$\dot u_0(t)\equiv0$, equation \eqref{eq:pf:Vdot_exact} and the nonlinear
bound \eqref{eq:pf:nonlinear_bound} give
\begin{equation}\label{eq:pf:Vdot_nominal}
    \dot V
    \le
    -\alpha_0\|\Bar e\|^2.
\end{equation}
Therefore, $V(t)\le V(0)$ for every $t\ge0$. Since \(V\) is positive
definite and radially unbounded in
$(\Bar e,\Phi,\Psi,\Bar r)$, it follows that
$\Bar e,\Phi,\Psi,\Bar r \in L_\infty.$
In particular, boundedness of $\Bar r$ and the constant source input
imply that the local adaptive input $\Bar u=\Bar r+\Bar u_0$ is also
bounded.

Integrating \eqref{eq:pf:Vdot_nominal} over $[0,T]$ gives
\begin{equation*}
    \alpha_0
    \int_0^T
    \|\Bar e(t)\|^2\,\mbox{d}t
    \le
    V(0)-V(T)
    \le
    V(0).
\end{equation*}
Letting $T\to\infty$ yields
\begin{equation}\label{eq:pf:e_L2}
    \Bar e\in L_2,
    \qquad
    \int_0^\infty
    \|\Bar e(t)\|^2\,\mbox{d}t
    <
    \infty.
\end{equation}

It remains to establish uniform continuity of $\Bar e$. From
\eqref{eq:AdpId:z_tracking_relation},
\begin{equation}\label{eq:pf:z_relation}
    \Bar z
    =
    \Bar x_0
    +
    (\mathbb A_m\mathbb L^{-1}\otimes I_n)\Bar e.
\end{equation}
Since $\Bar e\in L_\infty$ and $x_0\in L_\infty$ under the stated
bounded-source condition, \eqref{eq:pf:z_relation} implies
$\Bar z\in L_\infty.$

Moreover, global Lipschitz continuity of $g$ gives
\begin{align}
    \|\mathbf g(\Bar z)\|
    &\le
    L_g\|\Bar z\|
    +
    \sqrt m\|g(0)\|,
    \label{eq:pf:g_z_bounded}
\end{align}
and therefore $\mathbf g(\Bar z)\in L_\infty$.
Since $\Phi$, $\Psi$, $\Bar r$, $\Bar z$, and $\mathbf g(\Bar z)$
are bounded, the identity
$\Bar\eta=\Phi\Bar z+\Psi\mathbf g(\Bar z)+\mathbf B_0\Bar r$
implies that $\Bar\eta\in L_\infty$.
The bound \eqref{eq:pf:g_bound} and the boundedness of $\Bar e$
further imply that $\Delta\mathbf g\in L_\infty$.
It then follows from the nominal form of \eqref{eq:pf:error_dynamics}
that $\dot{\Bar e}\in L_\infty$.
Thus, $\Bar e$ is uniformly continuous. Since
$\Bar e\in L_2$, Barbalat's lemma gives
\begin{equation}\label{eq:pf:e_convergence}
    \lim_{t\to\infty}\|\Bar e(t)\|
    =
    0.
\end{equation}

Using $\Bar\epsilon=(\mathbb L^{-1}\otimes I_n)\Bar e,$
we consequently obtain
\begin{equation}\label{eq:pf:epsilon_convergence}
    \lim_{t\to\infty}\|\Bar\epsilon(t)\|
    =
    0.
\end{equation}

Finally, $\zeta=\Omega^\top P_{mn}\Bar e$ and
\eqref{eq:pf:e_convergence} imply
\begin{equation*}
    \lim_{t\to\infty}\|\zeta(t)\|
    =
    0.
\end{equation*}
For each $i\in\{1,\dots,m\}$, the adaptive laws satisfy
\begin{align*}
    \|\dot\Phi_i\|_F
    &\le
    \gamma_{A_i}\|\zeta_i\|\|z_i\|,\\
    \|\dot\Psi_i\|_F
    &\le
    \gamma_{C_i}\|\zeta_i\|\|g(z_i)\|.
\end{align*}
Since $z_i$ and $g(z_i)$ are bounded and $\zeta_i\to0$, it follows that
\begin{equation}\label{eq:pf:parameter_rate_convergence}
    \lim_{t\to\infty}\|\dot\Phi(t)\|_F
    =
    \lim_{t\to\infty}\|\dot\Psi(t)\|_F
    =
    0.
\end{equation}
Furthermore, under $\dot u_0\equiv0$,
\eqref{eq:pf:r_dynamics_global} reduces to
\begin{equation*}
    \dot{\Bar r}
    =
    -\Gamma_u\mathbf B_0^\top\zeta.
\end{equation*}
Therefore,
\begin{equation}\label{eq:pf:r_rate_convergence}
    \lim_{t\to\infty}\|\dot{\Bar r}(t)\|
    =
    0.
\end{equation}
This proves item~(ii).
\hfill$\square$

\paragraph*{Proof of Corollary~\ref{cor:AdpId:ISS}:}
Consider the error Lyapunov function
\begin{equation}\label{eq:pf:ISS_W}
    W
    \coloneqq
    \Bar e^\top P_{mn}\Bar e.
\end{equation}
Along the complete error dynamics \eqref{eq:AdpId:errordyn}, its
derivative satisfies
\begin{align}
    \dot W
    &=
    -\Bar e^\top Q_{mn}\Bar e
    +
    2\zeta^\top\Bar\eta
    +
    2\zeta^\top\mathbf C_0\Delta\mathbf g
    -
    2\zeta^\top\Bar\delta_0,
    \label{eq:pf:ISS_Wdot_raw}
\end{align}
where
$\zeta=\Omega^\top P_{mn}\Bar e$.
Define
\begin{equation*}
    c_\zeta
    \coloneqq
    \|\Omega^\top P_{mn}\|_2,
\end{equation*}
and recall from \eqref{eq:pf:alpha_0} that
\begin{equation*}
    \alpha_0
    =
    \lambda_{\min}(Q_{mn})
    -
    2c_\zeta
    \|C_0\|_2
    L_g
    \|\mathbb A_m\mathbb L^{-1}\otimes I_n\|_2
    >
    0.
\end{equation*}
Using the nonlinear mismatch bound
\eqref{eq:AdpId:g_mismatch_bound}, we obtain
\begin{equation}\label{eq:pf:ISS_Wdot_bound1}
    \dot W
    \le
    -\alpha_0\|\Bar e\|^2
    +
    2c_\zeta\|\Bar e\|\|\Bar\eta\|
    +
    2c_\zeta\|\Bar e\|\|\Bar\delta_0\|.
\end{equation}
Applying Young's inequality to the last two terms gives
\begin{subequations}
\begin{align}
    2c_\zeta\|\Bar e\|\|\Bar\eta\|
    &\le
    \frac{\alpha_0}{4}\|\Bar e\|^2
    +
    \frac{4c_\zeta^2}{\alpha_0}\|\Bar\eta\|^2,
    \label{eq:pf:ISS_eta_young}\\
    2c_\zeta\|\Bar e\|\|\Bar\delta_0\|
    &\le
    \frac{\alpha_0}{4}\|\Bar e\|^2
    +
    \frac{4c_\zeta^2}{\alpha_0}\|\Bar\delta_0\|^2.
    \label{eq:pf:ISS_delta_young}
\end{align}
\end{subequations}
Substituting these bounds into
\eqref{eq:pf:ISS_Wdot_bound1} yields
\begin{equation}\label{eq:pf:ISS_Wdot}
    \dot W
    \le
    -\frac{\alpha_0}{2}\|\Bar e\|^2
    +
    c_{\eta,W}\|\Bar\eta\|^2
    +
    c_{\delta,W}\|\Bar\delta_0\|^2,
\end{equation}
where
\begin{equation*}
    c_{\eta,W}
    =
    c_{\delta,W}
    \coloneqq
    \frac{4c_\zeta^2}{\alpha_0}.
\end{equation*}
Since $W\le\lambda_{\max}(P_{mn})\|\Bar e\|^2,$
we have
\begin{equation*}
    \|\Bar e\|^2
    \ge
    \frac{W}{\lambda_{\max}(P_{mn})}.
\end{equation*}
Consequently,
\begin{equation}\label{eq:pf:ISS_comparison}
    \dot W
    \le
    -\theta W
    +
    c_{\eta,W}\|\Bar\eta\|^2
    +
    c_{\delta,W}\|\Bar\delta_0\|^2,
\end{equation}
where
\begin{equation*}
    \theta
    \coloneqq
    \frac{\alpha_0}
    {2\lambda_{\max}(P_{mn})}
    >
    0.
\end{equation*}
Multiplying \eqref{eq:pf:ISS_comparison} by
$\mathrm e^{\theta t}$ and integrating over $[0,t]$ gives
\begin{align}
    W(t)
    \le{}&
    \mathrm e^{-\theta t}W(0)
    +
    c_{\eta,W}
    \int_0^t
    \mathrm e^{-\theta(t-\tau)}
    \|\Bar\eta(\tau)\|^2
    \,\mbox{d}\tau
    \nonumber\\
    &+
    c_{\delta,W}
    \int_0^t
    \mathrm e^{-\theta(t-\tau)}
    \|\Bar\delta_0(\tau)\|^2
    \,\mbox{d}\tau.
    \label{eq:pf:ISS_integral}
\end{align}
Define
\begin{equation*}
    \|\Bar\eta\|_{[0,t],\infty}
    \coloneqq
    \sup_{0\le\tau\le t}\|\Bar\eta(\tau)\|,
    \qquad
    \|\Bar\delta_0\|_{[0,t],\infty}
    \coloneqq
    \sup_{0\le\tau\le t}\|\Bar\delta_0(\tau)\|.
\end{equation*}
Since
\begin{equation*}
    \int_0^t
    \mathrm e^{-\theta(t-\tau)}
    \,\mbox{d}\tau
    =
    \frac{1-\mathrm e^{-\theta t}}{\theta}
    \le
    \frac{1}{\theta},
\end{equation*}
equation \eqref{eq:pf:ISS_integral} yields
\begin{align}
    W(t)
    \le{}&
    \mathrm e^{-\theta t}W(0)
    +
    \frac{c_{\eta,W}}{\theta}
    \|\Bar\eta\|_{[0,t],\infty}^2
    +
    \frac{c_{\delta,W}}{\theta}
    \|\Bar\delta_0\|_{[0,t],\infty}^2.
    \label{eq:pf:ISS_W_bound}
\end{align}
Using $\lambda_{\min}(P_{mn})\|\Bar e(t)\|^2\le W(t)$
in \eqref{eq:pf:ISS_W_bound} gives
\begin{align}
    \|\Bar e(t)\|^2
    \le{}&
    \frac{\mathrm e^{-\theta t}W(0)}
    {\lambda_{\min}(P_{mn})}
    +
    \frac{c_{\eta,W}}
    {\theta\lambda_{\min}(P_{mn})}
    \|\Bar\eta\|_{[0,t],\infty}^2 \nonumber \\
    &\quad+
    \frac{c_{\delta,W}}
    {\theta\lambda_{\min}(P_{mn})}
    \|\Bar\delta_0\|_{[0,t],\infty}^2.
    \label{eq:pf:ISS_e_squared_1}
\end{align}
Moreover, $W(0)\le\lambda_{\max}(P_{mn})\|\Bar e(0)\|^2.$
Therefore,
\begin{align}
    \|\Bar e(t)\|^2
    \le{}&
    \frac{\lambda_{\max}(P_{mn})}
    {\lambda_{\min}(P_{mn})}
    \mathrm e^{-\theta t}\|\Bar e(0)\|^2
    +
    \frac{c_{\eta,W}}
    {\theta\lambda_{\min}(P_{mn})}
    \|\Bar\eta\|_{[0,t],\infty}^2 \nonumber \\
    &\quad +
    \frac{c_{\delta,W}}
    {\theta\lambda_{\min}(P_{mn})}
    \|\Bar\delta_0\|_{[0,t],\infty}^2.
    \label{eq:pf:ISS_e_squared_2}
\end{align}
Taking the square root of both sides of
\eqref{eq:pf:ISS_e_squared_2} and using
$\sqrt{a+b+c}\le\sqrt a+\sqrt b+\sqrt c$ and
$\sqrt{\mathrm e^{-\theta t}}=\mathrm e^{-\theta t/2}$, we obtain
\begin{align}
    \|\Bar e(t)\|
    \le{}&
    \sqrt{
        \frac{\lambda_{\max}(P_{mn})}
        {\lambda_{\min}(P_{mn})}
    }\,
    \mathrm e^{-\theta t/2}\|\Bar e(0)\|
    +
    \sqrt{
        \frac{c_{\eta,W}}
        {\theta\lambda_{\min}(P_{mn})}
    }\,
    \|\Bar\eta\|_{[0,t],\infty}
    \nonumber\\
    &\quad+
    \sqrt{
        \frac{c_{\delta,W}}
        {\theta\lambda_{\min}(P_{mn})}
    }\,
    \|\Bar\delta_0\|_{[0,t],\infty}.
    \label{eq:pf:ISS_e_bound}
\end{align}

Thus, defining
\begin{equation*}
    \vartheta_{e,0}
    \coloneqq
    \sqrt{
        \frac{\lambda_{\max}(P_{mn})}
        {\lambda_{\min}(P_{mn})}
    },\qquad
    \vartheta_{e,\eta}
    \coloneqq
    \sqrt{
        \frac{c_{\eta,W}}
        {\theta\lambda_{\min}(P_{mn})}
    },\qquad
    \vartheta_{e,\delta}
    \coloneqq
    \sqrt{
        \frac{c_{\delta,W}}
        {\theta\lambda_{\min}(P_{mn})}
    },
\end{equation*}
we obtain \eqref{eq:AdpId:ISS_finite_e}.

Next, from
$\Bar\epsilon=(\mathbb L^{-1}\otimes I_n)\Bar e,$
it follows that
\begin{equation*}
    \|\Bar\epsilon(t)\|
    \le
    \|\mathbb L^{-1}\otimes I_n\|_2
    \|\Bar e(t)\|.
\end{equation*}
Hence, \eqref{eq:AdpId:ISS_finite_epsilon} holds with
\begin{equation*}
\begin{aligned}
    \vartheta_{\epsilon,0}
    &\coloneqq
    \|\mathbb L^{-1}\otimes I_n\|_2
    \vartheta_{e,0},\\
    \vartheta_{\epsilon,\eta}
    &\coloneqq
    \|\mathbb L^{-1}\otimes I_n\|_2
    \vartheta_{e,\eta},\\
    \vartheta_{\epsilon,\delta}
    &\coloneqq
    \|\mathbb L^{-1}\otimes I_n\|_2
    \vartheta_{e,\delta}.
\end{aligned}
\end{equation*}

Finally, the assumed bounds imply
\begin{equation*}
    \|\Bar\eta\|_{[0,t],\infty}
    \le
    \eta^\ast,
    \qquad
    \|\Bar\delta_0\|_{[0,t],\infty}
    \le
    \delta^\ast.
\end{equation*}
Letting $t\to\infty$ in
\eqref{eq:AdpId:ISS_finite_e} and
\eqref{eq:AdpId:ISS_finite_epsilon}, and using
$\mathrm e^{-\lambda \theta/2}\to0$, yields
\begin{align*}
    \limsup_{t\to\infty}\|\Bar e(t)\|
    &\le
    \vartheta_{e,\eta}\eta^\ast
    +
    \vartheta_{e,\delta}\delta^\ast,\\
    \limsup_{t\to\infty}\|\Bar\epsilon(t)\|
    &\le
    \vartheta_{\epsilon,\eta}\eta^\ast
    +
    \vartheta_{\epsilon,\delta}\delta^\ast.
\end{align*}
This establishes \eqref{eq:AdpId:ISS_finite} and
\eqref{eq:AdpId:ISS}.
\hfill$\square$

\end{document}